\documentclass[prc,twocolumn,showpacs,preprintnumbers,amsmath,amssymb]{revtex4}

\usepackage{amsmath,amssymb,multirow,epsfig,bm,color}

\newcommand{\beq}{\begin{equation}}
\newcommand{\eeq}{\end{equation}}
\newcommand{\bea}{\begin{eqnarray}}
\newcommand{\eea}{\end{eqnarray}}

\newcommand{\benn}{\begin{displaymath}}
\newcommand{\eenn}{\end{displaymath}}

\begin{document}

\title{Thermoelectric effects in silicene nanoribbons}

\author{K. Zberecki$^1$, M. Wierzbicki$^1$, J. Barna\'s$^2$, R. Swirkowicz$^1$}
\affiliation{$^1$Faculty of Physics, Warsaw University of Technology, ul. Koszykowa 75, 00-662 Warsaw, Poland}
\affiliation{$^2$Faculty of Physics, Adam Mickiewicz University, ul. Umultowska 85, 61-614 Pozna\'n, Poland\\ and
 Institute of Molecular Physics, Polish Academy of Sciences, Smoluchowskiego 17, 60-179 Pozna\'n, Poland}
\date{\today}

\begin{abstract}
Transport and thermoelectric coefficients (including also spin thermopower)  of silicene nanoribbons with zigzag edges are
investigated by {\it ab-initio}  numerical methods. Local spin density of such nanoribbons reveals edge magnetism. Like in graphene, one finds antiferromagnetic and ferromagnetic ordering, with spin polarization
at one edge  antiparallel or parallel to that at the other edge, respectively. Thermoelectric properties, especially the Seebeck
coefficient, significantly depend on the electronic band structure and are enhanced when the Fermi level is in the energy gap.
However, these thermoelectric properties are significantly reduced when the phonon contribution to the heat conductance is included.
This phonon contribution has been calculated numerically by two different methods.
Transition from antiferromagnetic to ferromagnetic states leads to a large magnetoresistance as well as to a considerable magnetothermopower.
Thermoelectric parameters in the antiparallel configuration, when spin polarization in the left part of the nanoribbon
is opposite to that in the right part, are also analyzed.
\end{abstract}

\pacs{xxx}

\maketitle

\section{Introduction}

There is currently an increasing interest in two-dimensional conducting materials,
like graphene -- a two-dimensional hexagonal lattice of carbon atoms. Graphene exhibits
unusual transport properties which follow from its peculiar electronic structure. More
specifically, the low-energy electronic states around the Fermi level (K points of the
Brillouin zone) are described by the Dirac model and the electrons behave like massless particles.
Owing to its promising electronic properties,
like high electron mobility and long spin diffusion length, graphene is considered as an
ideal material for future nanoelectronic and spintronic devices~\cite{dlubak}. Therefore, not only
transport, but also magnetic and thermoelectric properties of graphene are currently of
great interest. Some possible applications of this novel material have already
been proposed~\cite{novoselov,berger}.

Very recently another two-dimensional material, silicene, has been fabricated~\cite{aufray,padova,vogt,kara}.
Silicene is a two-dimensional hexagonal lattice of silicon atoms, but contrary to graphene, silicene
has a buckled atomic structure  -- the two triangular sublattices are slightly displaced
vertically. Electronic structure of the two-dimensional silicene is similar to that of graphene,
i.e. silicene is a semimetal with low-energy states at the Fermi level described by the
Dirac model. Spin-orbit interaction opens an energy gap at the Fermi level, but this gap,
like in graphene,  is rather small. From the application point of view, however, semiconducting
transport properties are more desired than metallic ones. Thus, opening a gap at the Fermi
level (K points) in the electronic spectrum is one of key challenges. One way to achieve
this objective is to form quasi-one-dimensional nanoribbons. Indeed, fabrication of such
silicene nanoribbons (SiNRs) has been reported recently~\cite{padova,vogt}, which opened
new perspectives for this novel material~\cite{kara,cahagirov}. Therefore, detailed description
and understanding of physical properties of silicene is currently of great interest.

Electronic, mechanical and magnetic properties of SiNRs have been studied recently by
first-principle numerical methods~\cite{cahagirov,xu,kang,pan}. In particular, electronic
transport properties of  SiNRs with zigzag edges (zSiNRs) have revealed a magnetoresistance
effect~\cite{xu} associated with transition of the edge magnetism from ferromagnetic to antiferromagnetic
ordering. In turn, the giant magnetoresistance effect can be observed in narrow
ribbons with zigzag chains, when magnetizations of the external electrodes change from
antiparallel to parallel alignment, for instance in an external magnetic field~\cite{kang}.
Some preliminary calculations of thermoelectric properties of armchair as well as zigzag
SiNRs have also been reported~\cite{pan}. Relatively high thermopower $S$ and some enhancement
of the thermoelectric efficiency have been found at high temperatures for nonmagnetic armchair
ribbons of some specific widths. Results obtained for zSiNRs have revealed less remarkable effects.

Thermoelectric properties of nanoscopic systems are currently of great interest due to the
possibility of heat to electrical energy conversion at nanoscale, which is important for applications. 
Quantum confinement and transport blockade can lead to a considerable
enhancement of the thermoelectric efficiency in such structures~\cite{venkata,hochbaum,harman,duarte}.
An interplay between the spin effects and thermoelectric properties in magnetic tunnel junctions
and nanoscale systems has been also intensively studied in view of possible applications
in spintronic devices~\cite{walter,liebing,uchida,wierzbicki}. As a result, some new spin-related thermoelectric
phenomena have been discovered. Certainly, the most spectacular spin-related
effect is the spin thermopower (spin Seebeck effect), which is a spin analog of the usual electrical
thermopower (Seebeck effect)~\cite{uchida}. As the conventional thermopower consists in generation
of electrical voltage in an open system by a temperature gradient, the spin thermopower corresponds
to the thermal generation of spin voltage.

In this paper we analyze thermoelectric properties of SiNRs with zigzag edges. The calculations
have been carried out by ab-initio numerical methods. Narrow Si nanoribbons, similarly
to graphene ones~\cite{son,pisani,rojas}, reveal antiferromagnetic (AFM) ordering -- the spin polarization
at one edge is opposite to that at the other edge. Moreover, the corresponding electronic
spectrum has an energy gap in the close vicinity of the Dirac points~\cite{pan,ding}. The AFM ordering
is shown to have  a strong influence on the transport properties, especially on the thermopower $S$, which can be considerably
enhanced in systems with relatively wide gaps. Therefore, the accurate determination of the gap
is crucial for the proper description of thermoelectric properties. 

By applying an external
magnetic field, one can switch the magnetic configuration from AFM to ferromagnetic (FM) one. 
This change, in turn, leads to significant changes in transport properties. Accordingly, there is a large magnetoresistance
associated with transition from the AFM state to the FM one. As we show in this paper, the thermopower
also strongly depends on the magnetic configuration, so a considerable magnetothermopower can be
observed. The latter effect is much stronger than in standard magnetic tunnel junctions~\cite{walter,liebing}.
Therefore, Si nanoribbons could be considered as interesting systems for applications
in future nanoelectronics.

The paper is organized as follows. In Sec. 2 we describe the computational method used to
determine transmission through the system. This transmission is subsequently used to determine
the thermoelectric coefficients. The obtained numerical results on electronic transport and on
electronic contribution to the heat transport in the limit of spin channel mixing are presented
and discussed in Sec. 3. In turn, spin thermoelectric properties are considered in Sec. 4.
Heat transport mediated by phonons is presented and discussed in Sec. 5.
Summary and concluding remarks are presented in Sec. 6.

\section{Computational details: spin density and transmission function}

\begin{figure*}[ht]
  \begin{center}
    \begin{tabular}{cc}
      \resizebox{70mm}{!}{\includegraphics[angle=0]{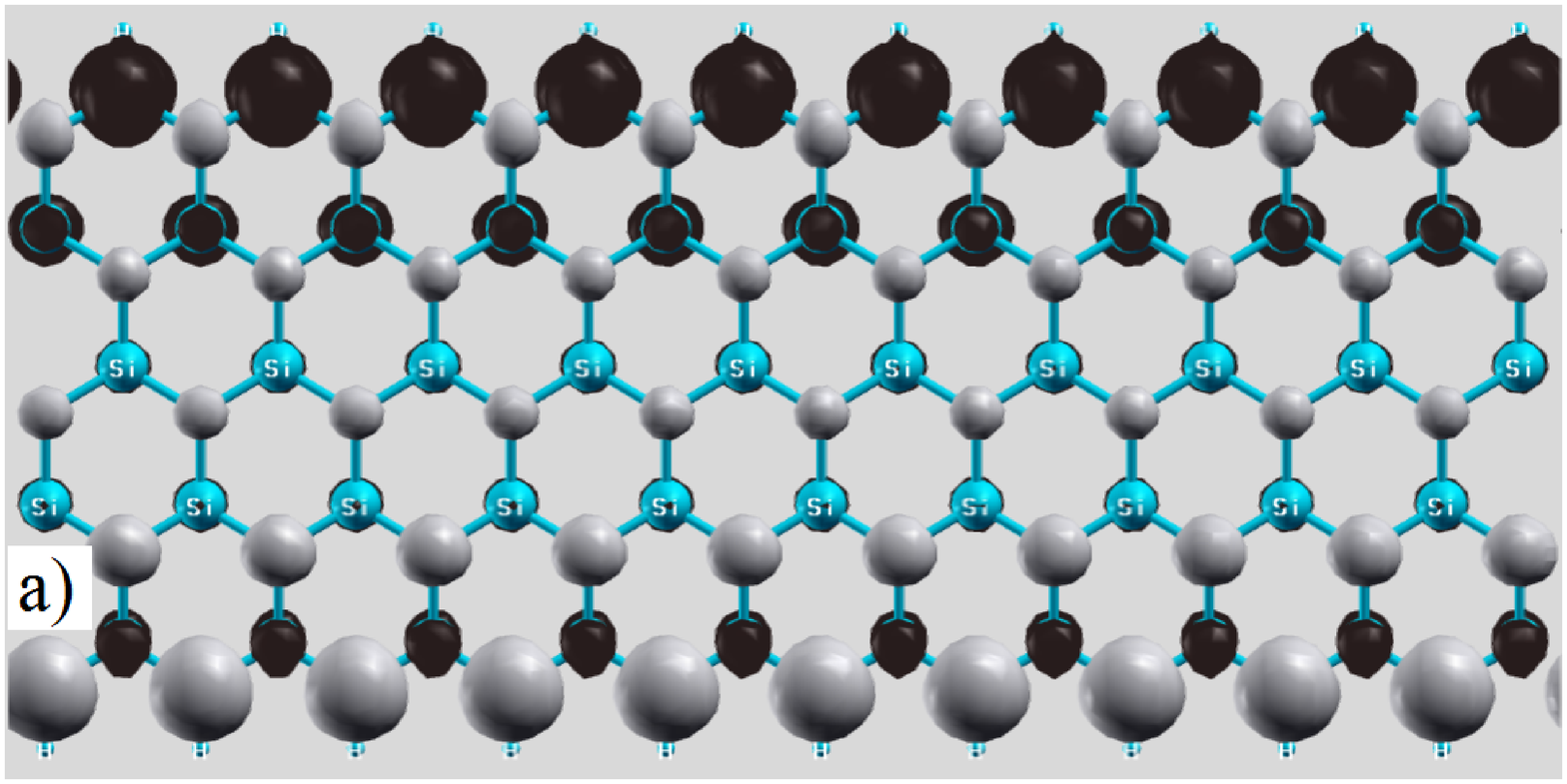}} &
      \resizebox{70mm}{!}{\includegraphics[angle=0]{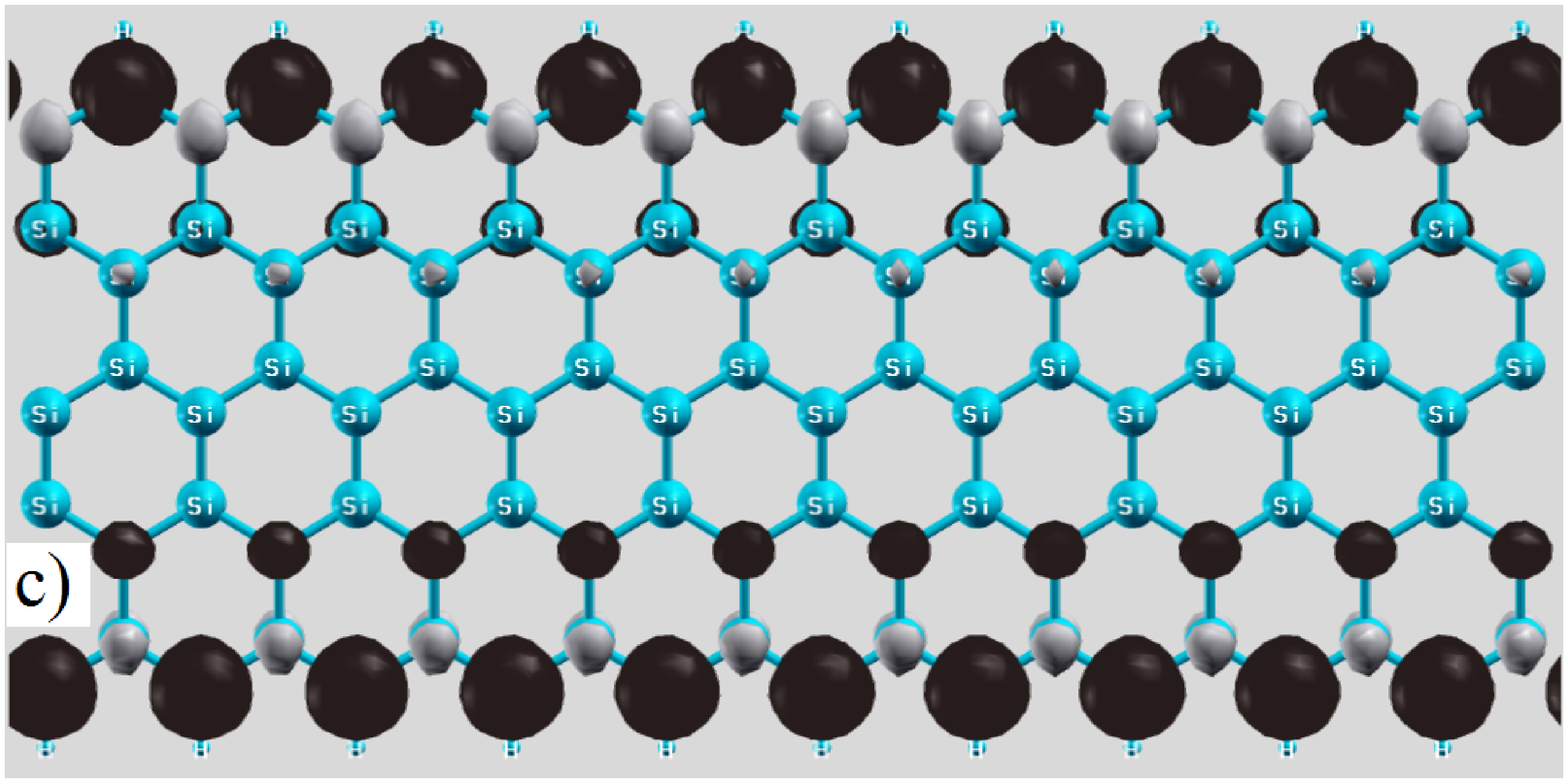}} \\
      \resizebox{80mm}{!}{\includegraphics[angle=270]{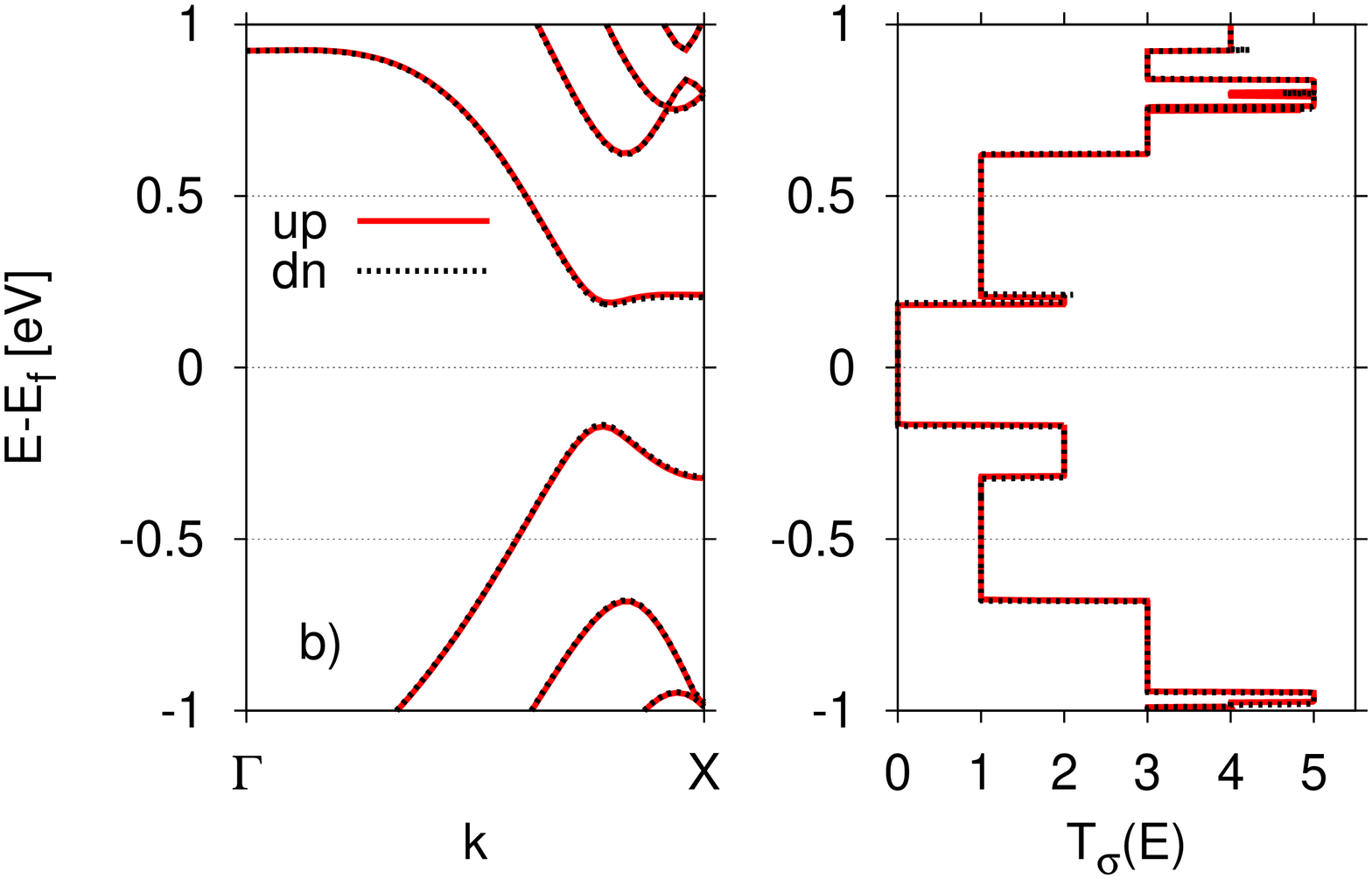}} &
      \resizebox{80mm}{!}{\includegraphics[angle=270]{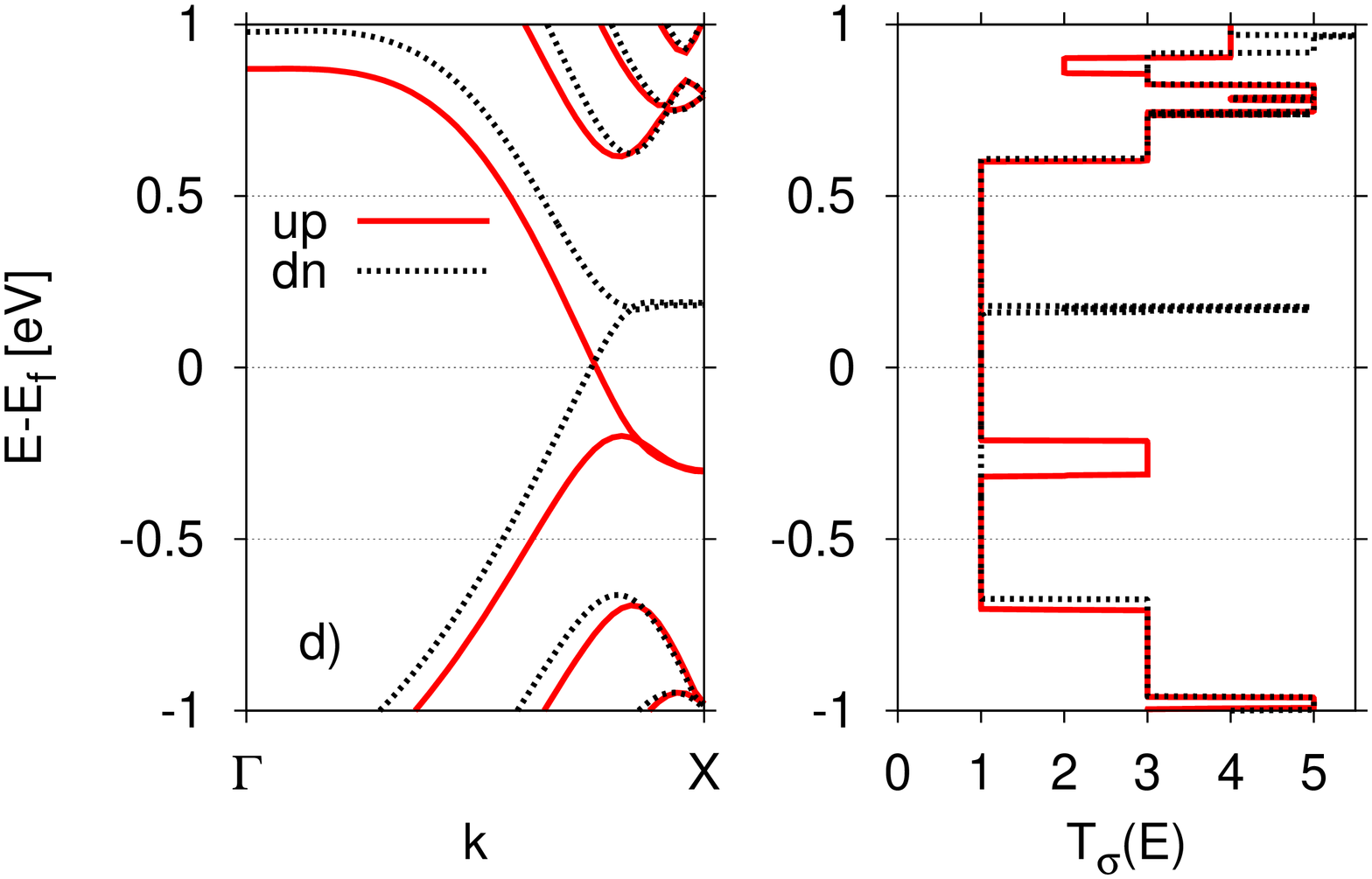}} \\
    \end{tabular}
    \caption{Spin density in the AFM (a) and FM (c) states, calculated within GGA approximation for zSiNRs with $N=5$.
             The corresponding spin-resolved band structure and  transmission function are shown
             in (b) and (d) for the AFM and FM configurations, respectively. The energy is measured with respect to the corresponding Fermi energy $E_f$  of an undoped structure (note $E_f$ for AFM and FM states are generally  different). }
    \label{fig1}
  \end{center}
\end{figure*}

Electronic transport through zSiNRs was investigated numerically by ab-initio calculations within
the DFT Siesta code~\cite{siesta1}. The spin-resolved energy-dependent transmission, $T_{\sigma}(E)$, through
zSiNRs of different widths was  determined in terms of the non-equilibrium Green function
method (NGF) as implemented in the Transiesta code~\cite{siesta2}. As in Ref.~\cite{son1},
width of a zSiNR is characterized by the corresponding  number $N$ of zigzag chains in the ribbon.
The nanoribbon edges were terminated with hydrogen atoms to remove the dangling bonds.
The structures were optimized until atomic forces converge to 0.02 eV/A. The atomic
double-$\zeta$ polarized basis (DZP) was used and the grid mesh cutoff was set equal to 200 Ry.
The generalized gradient approximation (GGA) with Perdrew-Burke-Ernzerhof parameterization
was applied for exchange-correlation part of the total energy functional~\cite{ca1}. For comparison,
some calculations have been also performed within the local density approximation (LDA) with
Ceperley-Alder parameterization (equivalent to Perdew-Zunger one)~\cite{pz1,pbe1}. The performed calculations,
similarly to those presented in Refs~\cite{pan,ding}, show that the AFM ordering is the most stable configuration in
narrow zSiNRs. Spin configuration in both AFM and FM states is shown in Figs \ref{fig1}(a) and \ref{fig1}(c), respectively.
The energy difference between the
FM and AFM configurations  is rather small and for
$N=5$ it is equal to 0.02 eV. Thus, the configuration can be easily changed from
AFM to FM one by applying an external magnetic field. Therefore, the following calculations have been
performed for both AFM and FM magnetic states.

Spin-polarized band structure and the corresponding spin-dependent transmission functions
$T_{\sigma}(E)$ are presented in Fig. \ref{fig1}(b) and \ref{fig1}(d) for the AFM and FM configuration, respectively.
Both band structure and transmission functions are significantly different in the  two magnetic states.
In the AFM case, the bands are spin degenerate, and an energy gap exists in the vicinity of $E=0$
(corresponding to the Dirac points). Numerical calculations clearly show that the gap width decreases
with increasing $N$. On the other hand, the transmission function for the FM configuration is spin dependent
and is constant and finite near the Dirac points for both spin orientations (indicating absence of energy gap) in agreement with Refs~\cite{xu,ding}. These results are also
qualitatively similar to those obtained for graphene nanoribbons~\cite{rojas}.

\section{Thermoelectric properties}

In this section we present numerical results on conventional thermoelectric
properties of zSiNRs for $N$=5, 6, and 7. By conventional thermoelectricity
we mean here the effects that occur when the two spin channels are mixed
in the contacts, so no spin thermopower can be observed. Later on we will
come back to the problem of spin thermoelectric phenomena.

In the linear response regime, the electric $I$ and heat $I_{Q}$ currents
flowing through the system from left to right, when the electric
potential and temperature of the left electrode are higher by $\Delta V$  and  $\Delta T$,
respectively, can be written in the matrix form as

\beq
\left(
\begin{array}{c}
I   \\
I_{Q} \\
\end{array}
\right)
=
\left(
\begin{array}{cc}
e^2 L_{0} & \frac{e}{T}L_{1} \\
 eL_{1} &  \frac{1}{T}L_{2} \\
\end{array}
\right)
\left(
\begin{array}{c}
\Delta V   \\
\Delta T \\
\end{array}
\right),
\eeq
where $e$ is the electron charge, while $L_{n}$ ($n=0,1,2$) are defined as
$L_{n} = -\frac{1}{h} \int dE\,T(E)\, (E-\mu)^{n} \frac{\partial f}{\partial E}$,
with $f(E)$ being the equilibrium Fermi-Dirac distribution function corresponding
to the chemical potential $\mu$ and temperature $T$ (equal in both electrodes), and
$T(E) = \sum_{\sigma} T_{\sigma}(E)$ denoting the total transmission through the system
(maximum transmission is equal to the number of different quantum channels).
We have determined the transmission function $T(E)$ using the $ab-initio$ method,
as described in the preceding section. Having found $T(E)$, one can calculate the
functions $L_{n}$ by integrating over energy. Then, the electrical conductance $G$ is given by
$G=e^{2}L_{0}$, while the
electronic contribution to the thermal conductance, $\kappa_{e}$, is
\begin{equation}
\kappa_{e}=\frac{1}{T}(L_{2} - \frac{L_{1}^{2}}{L_{0}}).
\end{equation}
In turn,  the thermopower, $S=-\Delta V /\Delta T$, can be calculated from the formula~\cite{wierzbicki}
\begin{equation}
S=-\frac{L_{1}}{|e|TL_{0}}.
\end{equation}
Now, we consider the thermoelectric effects for some specific spin arrangements at the nanoribbon
edges.

\subsection{AFM configuration}

\begin{figure}[ht!]
    \begin{tabular}{c}
      \resizebox{80mm}{!}{\includegraphics[angle=270]{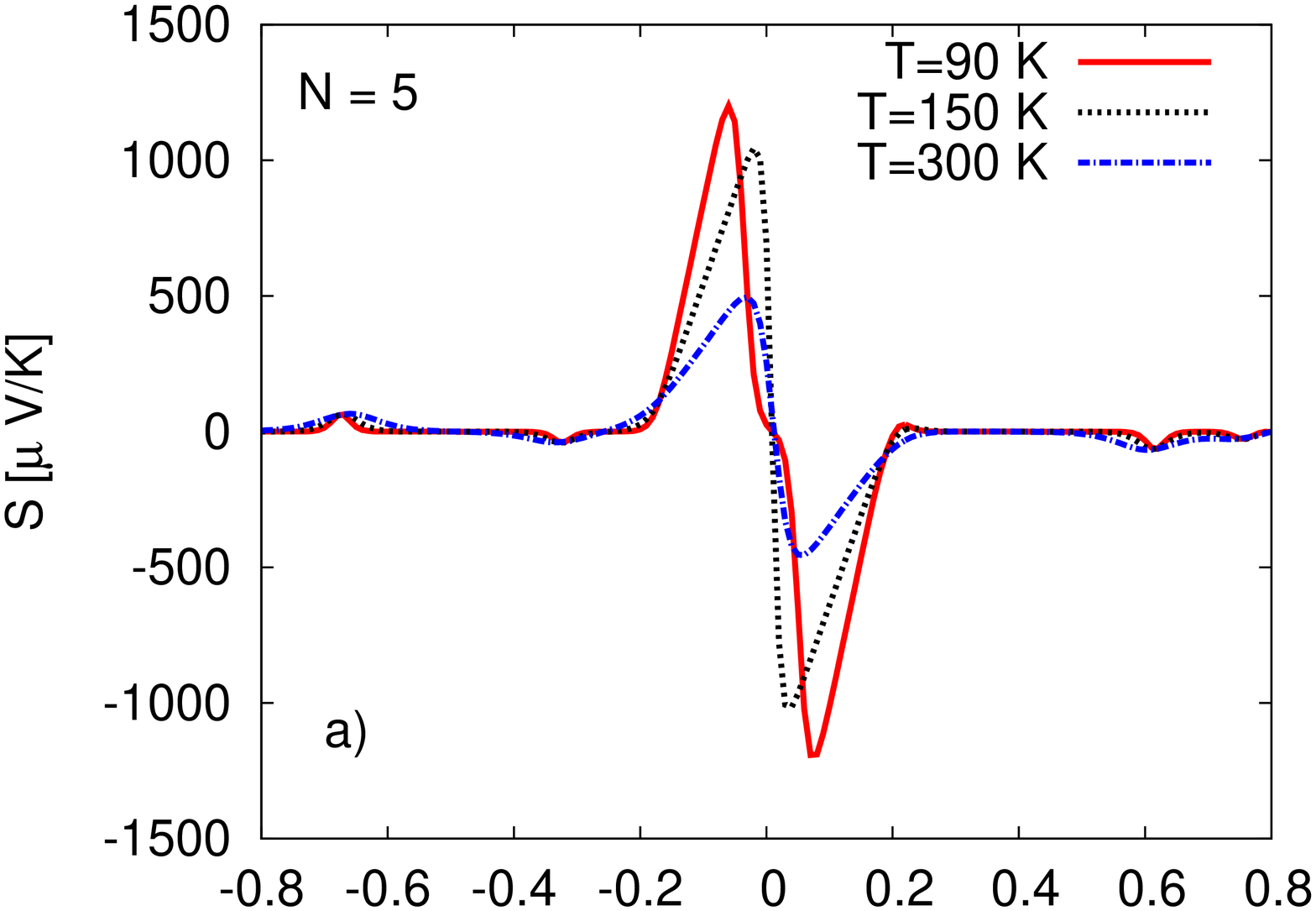}} \\
      \resizebox{80mm}{!}{\includegraphics[angle=270]{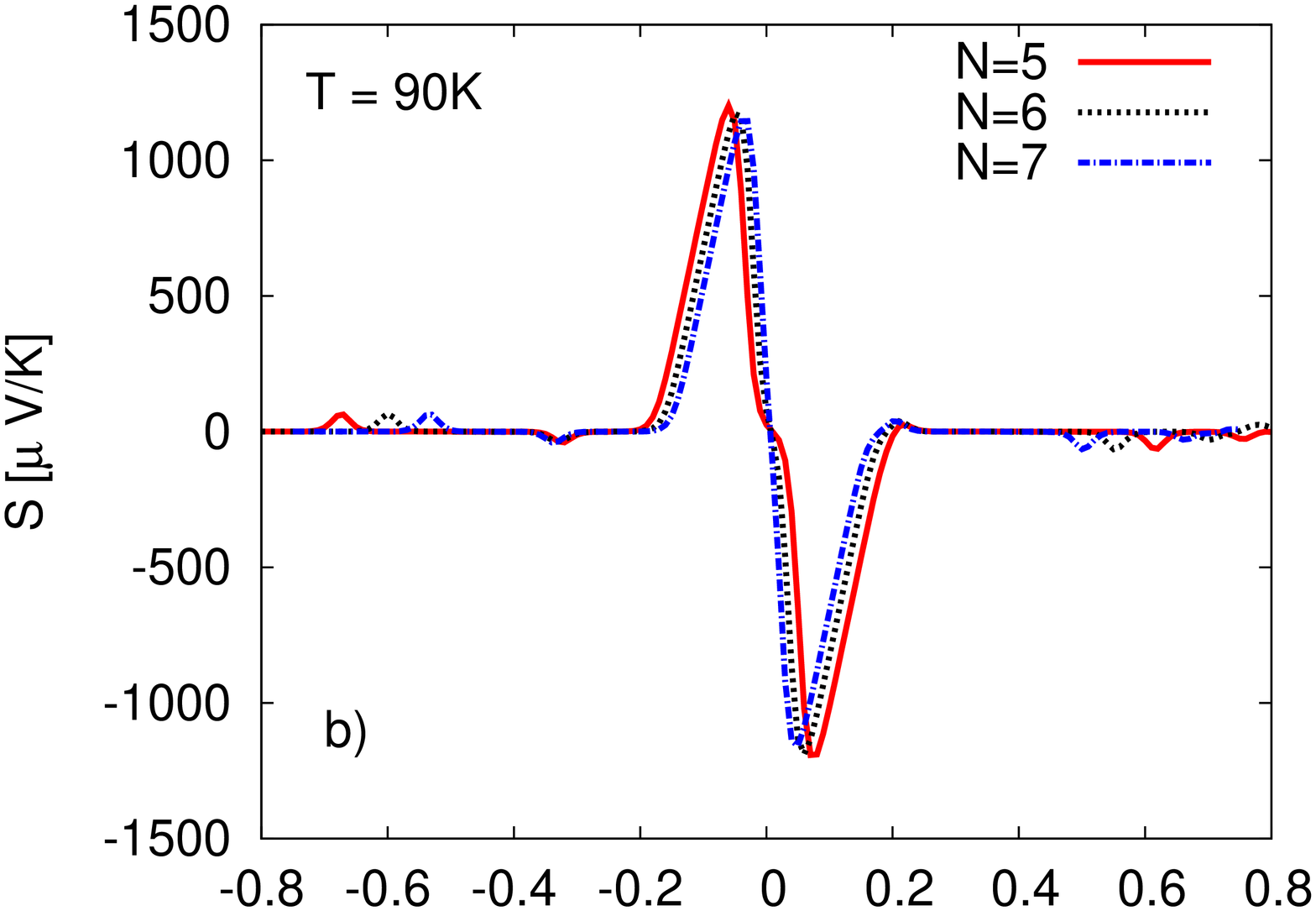}} \\
      \resizebox{80mm}{!}{\includegraphics[angle=270]{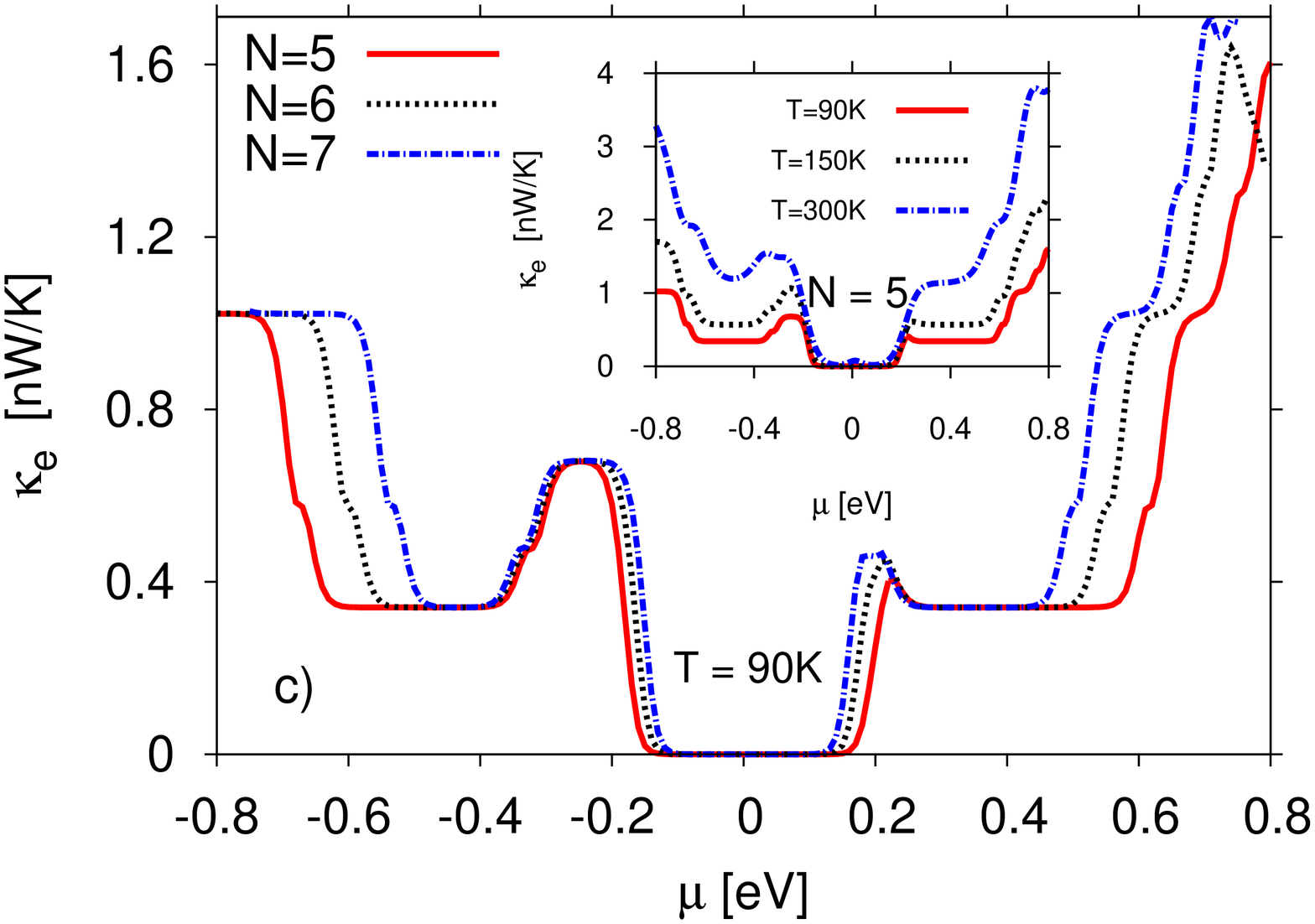}}
    \end{tabular}
    \caption{Thermopower $S$ (a,b) and electronic thermal conductance $\kappa_{e}$ (c) of zSiNRs in
             the AFM configuration  as a function of the chemical potential $\mu$, calculated within the GGA
             approximation for different values of $N$ and temperature, as indicated.}
    \label{fig2}
\end{figure}

The AFM configuration is the most stable configuration, at least in the regime
of low temperatures. Later we will consider other configurations, which correspond
to a higher total energy. Results obtained for the thermopower $S$ and thermal
conductance $\kappa_{e}$ in the AFM state are shown in Fig. \ref{fig2}, where $S$ and $\kappa_{e}$ are presented
as a function of the chemical potential $\mu$. In reality, the chemical potential $\mu$
can be changed in the vicinity of the Dirac points (corresponding to $\mu$=0) either
by p-type or n-type doping, which results in $\mu<0$ and $\mu>0$, respectively.
In general, the chemical potential could be also varied with an external gate voltage.

Figure \ref{fig1}(b) clearly shows the presence of a relatively wide energy gap in the
transmission function for the AFM configuration, which extends roughly from $-0.2$ to $+0.2$ eV.
This gap has a strong influence on the transport and thermoelectric  properties. First,
the zero-temperature electrical conductance vanishes for chemical potential in the gap.
As the temperature is sufficiently high, transport is mediated by activated electrons
and/or holes. Consider the thermopower shown in Fig. \ref{fig2}(a).
When $\mu=0$, the thermopower vanishes as
the currents {\it via} electrons and holes (though very small) compensate each other.
When $\mu$ is positive (but still inside the gap), the thermopower is negative and achieves
relatively large absolute values. The maximum of the absolute value of $S$ appears roughly
when $\mu$ is at the distance of the order of $kT$ from the upper edge of the gap. The hole current
is then blocked and the charge current is mediated by electrons. The thermocurrent flows then from right to left (electrons flow from left to right) and a positive voltage is needed to block the current. According to the definition, the thermopower is then negative.
When, in turn, the chemical potential is negative (and still inside the gap), the thermocurrent is dominated by holes and flows from left to right. To block the current one needs a negative
voltage and therefore the thermopower is positive.
Apart from the high peaks
inside the gap, the thermopower is rather small in the remaining part of the considered range
of $\mu$. This is due to partial (or total) compensation of the contributions to thermocurrent from electrons and holes. When $\mu$ is already in the conduction band and slightly above the narrow peak in the transmission seen in Fig.~\ref{fig1}(b), the thermopower becomes positive. For higher values of $\mu$ it becomes negative again. Generally, the sign of thermopower depends on the details of electronic spectrum.

The thermopower strongly depends on temperature. With increasing $T$ the intensities
of the main peaks are considerably reduced, although $S$ is enhanced in a narrow region of $|\mu|$
around $\mu=0$, see Fig. \ref{fig2}(a). Thus, quite remarkable values of $S$ can be achieved at high temperatures for very small
n- or p-type doping, despite of the energy gap in the transmission. Similar behavior of
the thermopower $S$ can be observed also in the nanoribbons with $N=6$ and $N=7$, see Fig.~2(b).
Some small differences follow from a decrease in the energy gap width
for wider nanoribbons.

In Fig. \ref{fig2}(c), the thermal conductance due to electron transport, $\kappa_{e}$,
is presented as a function of $\mu$. This figure clearly shows that the conductance  $\kappa_{e}$ is strongly
suppressed for $\mu$  in the energy gap. However, it
significantly increases when the chemical potential is outside the gap (in the regions of nonzero
transmission),  even at low temperatures. The range of $\mu$, where $\kappa_{e}$ is considerably
suppressed, depends on the nanoribbon width and also weakly on temperature (see the inset
in Fig \ref{fig2}(c)). It is interesting to note that for higher temperatures, a small maximum
develops in the vicinity of $\mu=0$. Such a maximum also appears in other systems~\cite{wierzbicki}.

\subsection{FM configuration}

\begin{figure}[ht]
    \begin{tabular}{c}
      \resizebox{80mm}{!}{\includegraphics[angle=270]{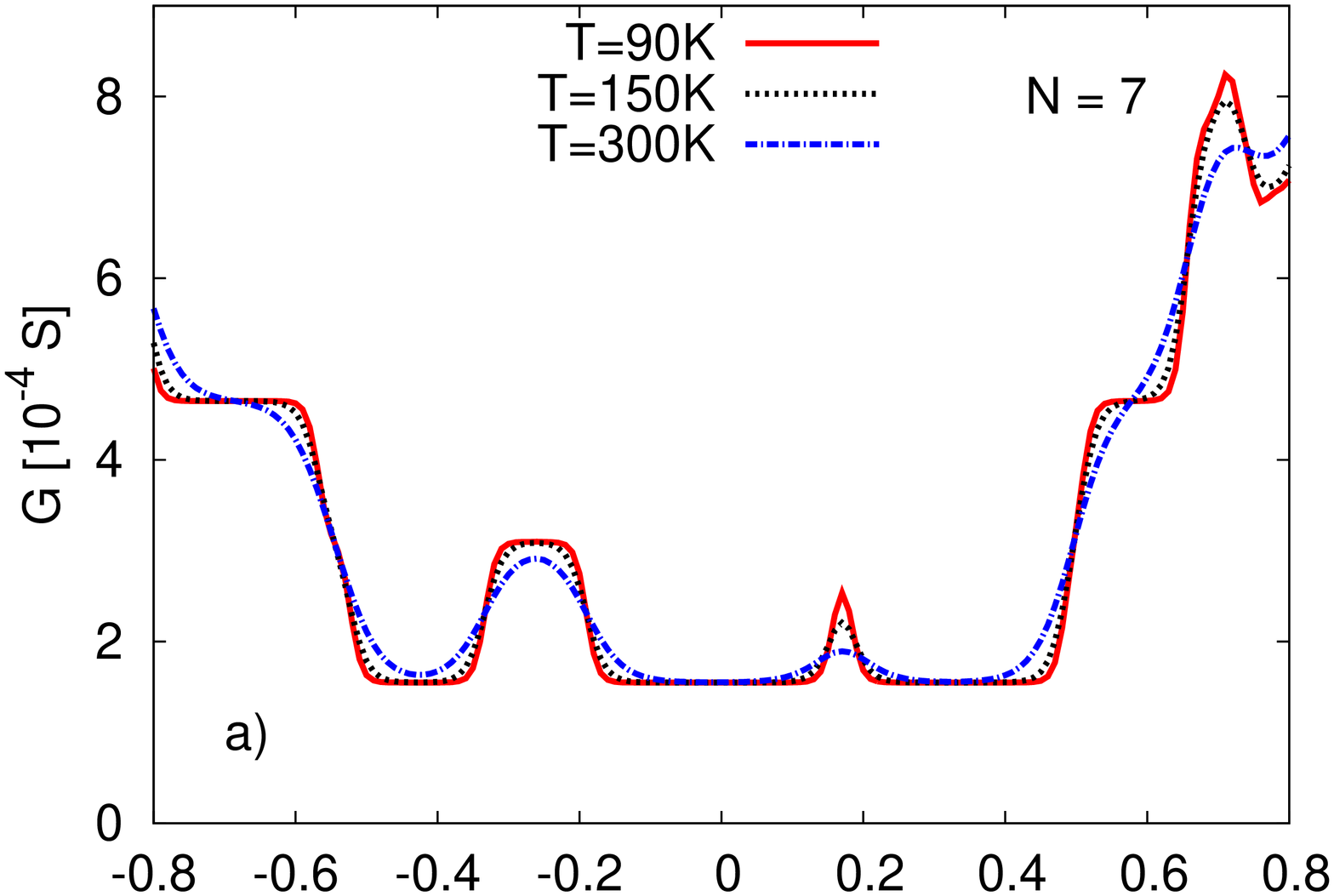}} \\
      \resizebox{80mm}{!}{\includegraphics[angle=270]{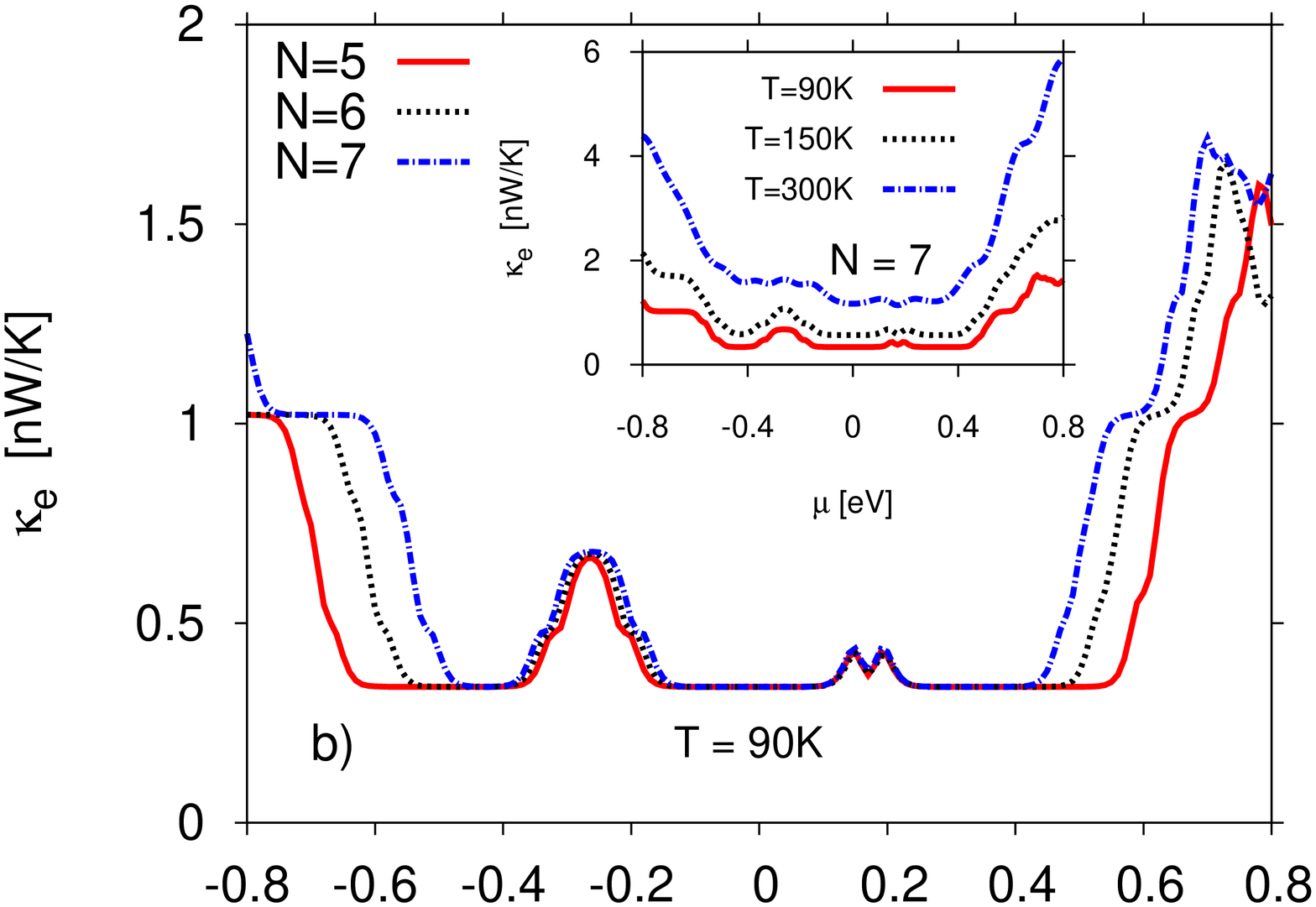}} \\
      \resizebox{80mm}{!}{\includegraphics[angle=270]{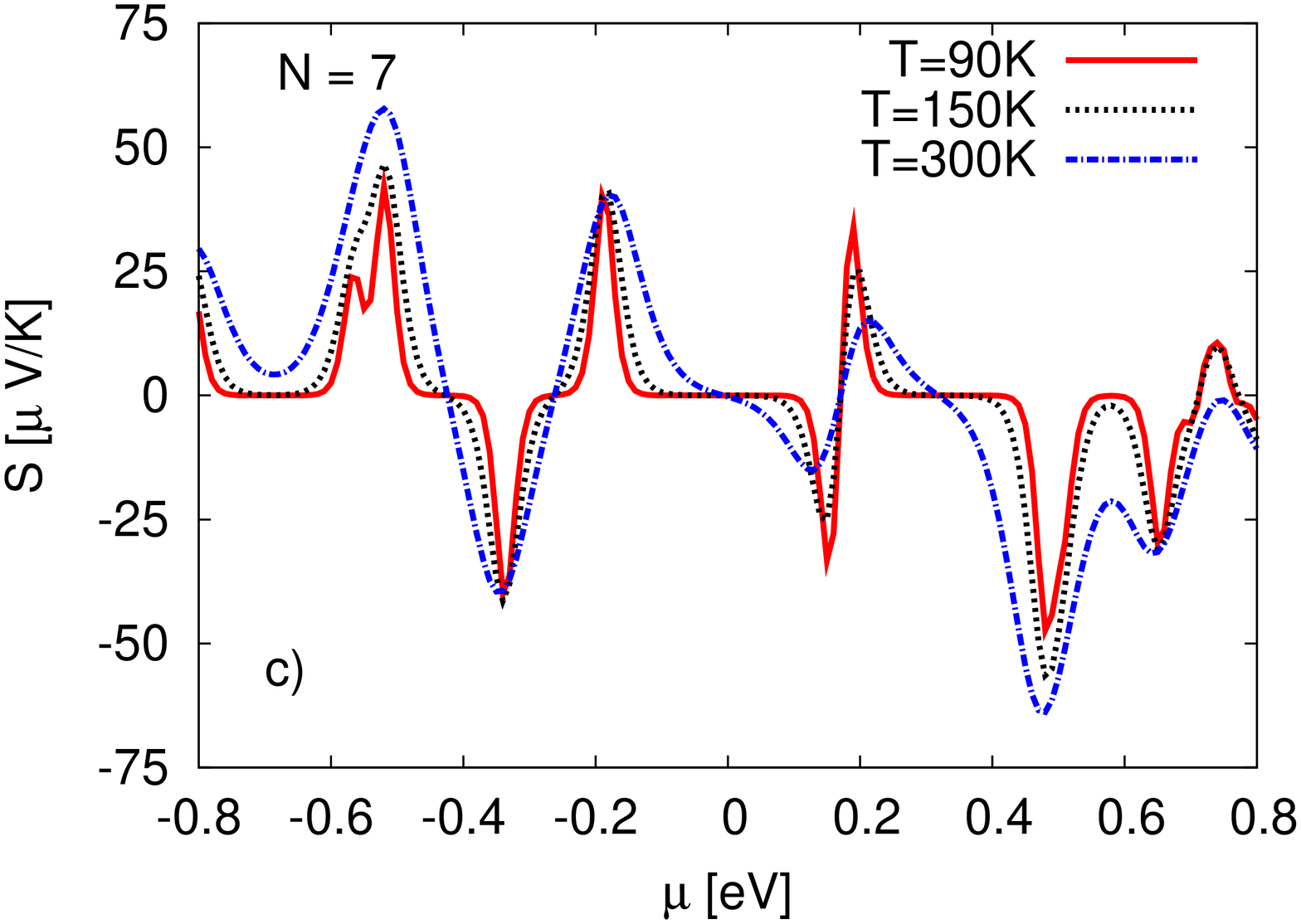}}
    \end{tabular}
    \caption{Electric conductance $G$ (a), electronic thermal conductance  $\kappa_{e}$ ( b ),
             and thermopower $S$ (c) in  the FM state, calculated as a function of chemical
             potential within GGA for given values of $N$ and $T$.}
    \label{fig3}
\end{figure}

Now, let us consider the ferromagnetic FM configuration. Although
 energy of this configuration is slightly higher than that of the AFM state, it
 can be stabilized by an external magnetic field. From Fig. \ref{fig1}(d) follows that
there is no energy gap in the FM configuration, and the system behaves like a metal with
a constant transmission function in a certain region of chemical potentials close to $\mu=0$.
Below that region (negative $\mu$) there
is a wide maximum in transmission for majority spins $\uparrow$, whereas above (positive $\mu$) a narrow peak
in transmission appears
for minority spins $\downarrow$. Apart from this, transmission only weakly depends on the nanoribbon
width. Numerical results on the transport parameters in the FM state are presented in Fig. \ref{fig3}.
The electrical conductance $G$ and the thermal conductance $\kappa_{e}$, calculated for several
temperatures, are shown in Figs \ref{fig3}(a) and \ref{fig3}(b), respectively. Both $G$ and $\kappa_{e}$ are
constant for small values of $\mu$, and display maxima for relatively high p and n-type doping,
which follow from the maxima in transmission function for majority and minority spins,
respectively.

\begin{figure}[ht]
  \begin{center}
    \begin{tabular}{cc}
      \resizebox{80mm}{!}{\includegraphics[angle=270]{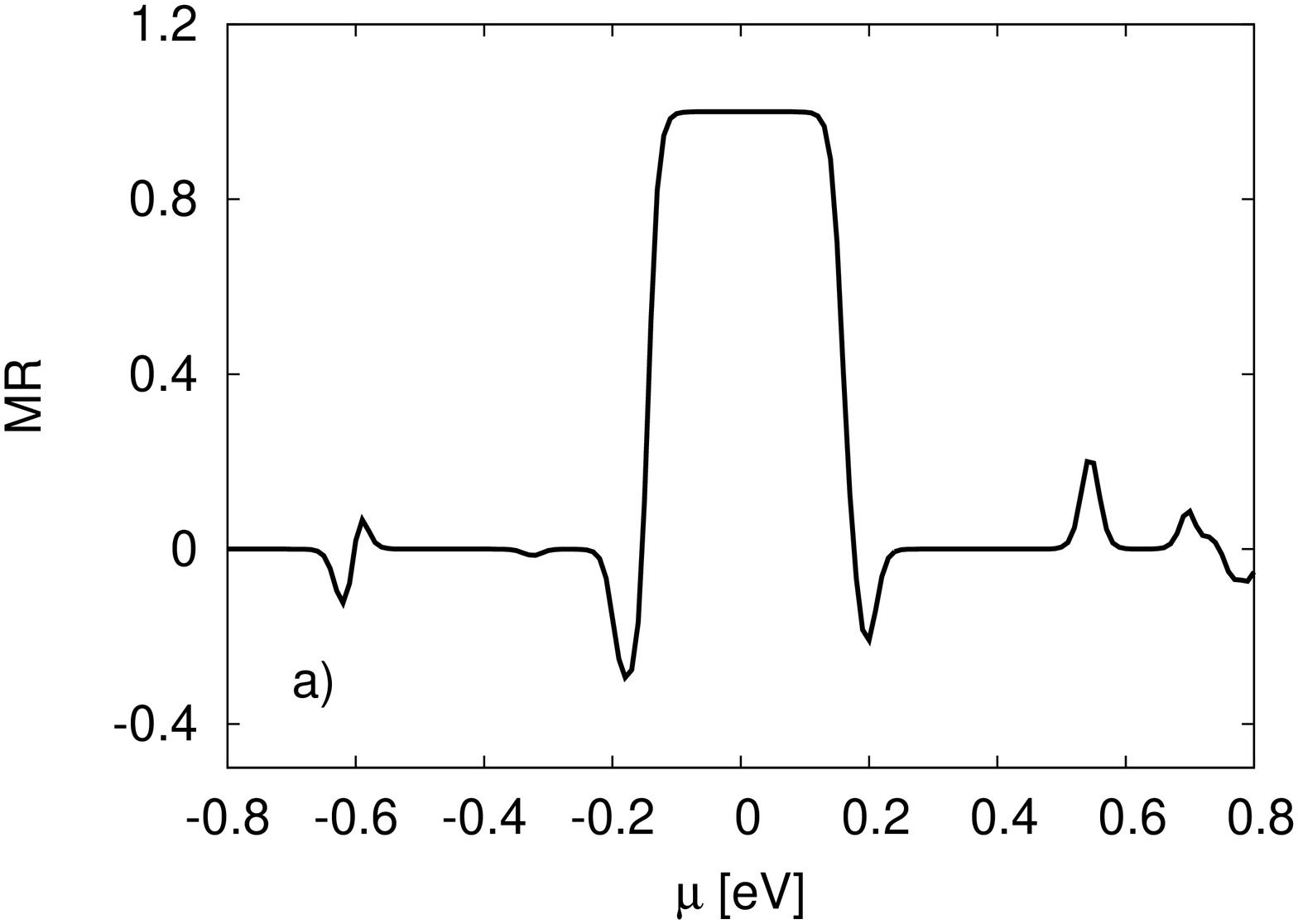}}\\ 
      \resizebox{80mm}{!}{\includegraphics[angle=270]{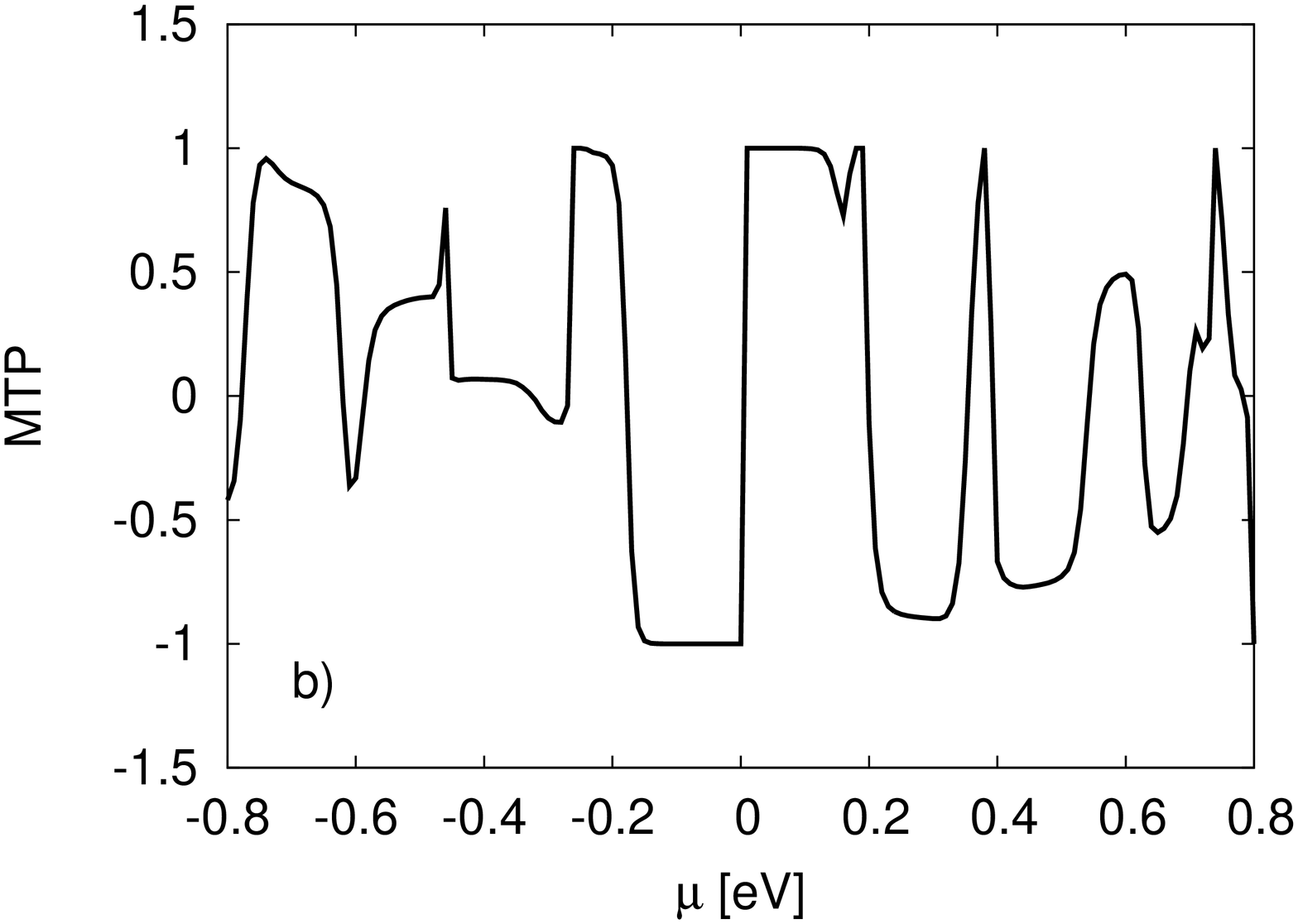}}
    \end{tabular}
    \caption{Magnetoresistance MR (a) and magnetothermopower MTP (b) associated with a change from the AFM to FM states as a function of $\mu$, calculated  within the GGA approximation
             for N=5 and T=90K}
    \label{fig4}
  \end{center}
\end{figure}

Due to the gap existing in the AFM state and its absence in the FM state, electrical and heat transport
are significantly different in these two configurations. More specifically, the conductance $G$ and heat conductance $\kappa_e$
at low doping are considerably larger in the FM state. Thus, one can observe a large magnetoresistance (MR) defined quantitatively
as ${\rm MR}=(G_{\rm FM}-G_{\rm AFM})/(G_{\rm FM}+G_{\rm AFM})$, where $G_{\rm AFM}$ and $G_{\rm FM}$
are the total electrical conductances in the AFM and FM states, respectively.
The magnetoresistance of nanoribbons corresponding
to $N=5$ is presented in Fig. \ref{fig4} as a function of the chemical potential. For small p- and n-type
doping, MR achieves values practically equal to 1, since $G_{\rm AFM}$ is close to zero due to the
energy gap. For higher doping, a negative magnetoresistance can be observed. These results
clearly show that narrow zSiNRs can be important elements for spintronic devices, in which
magnetic configuration can be easily changed from the AFM to FM states by an external magnetic
field.

The thermopower $S$ in the FM configuration, calculated for several values of
temperature, is presented in Fig. \ref{fig3}(c). When the chemical potential is changed, $S$ displays several
peaks with intensities weakly dependent on temperature. These peaks are now remarkably smaller than in the AFM configuration.
In particular, there is a region of small values of $\mu$, where $S$ is practically suppressed
to zero at
low temperatures, which is a consequence of a constant transmission in the FM configuration close to the Dirac points. This
region becomes narrower with increasing temperature, but $S$ is still very small. Comparing
the results presented in Figs \ref{fig2}(a) and 3(c), one can conclude that the thermopower remarkably
changes with magnetic configuration. To describe this dependence one can introduce the
magnetothermopower (MTP), defined quantitatively as ${\rm MTP} = (S_{\rm FM}-S_{\rm AFM}/(|S_{\rm FM}|+|S_{\rm AFM}|)$.
Such a definition is convenient as it allows avoiding artifacts in the regions, where the thermopowers in both configurations have
opposite signs and are similar in magnitude.
The calculated MTP, presented in Fig. \ref{fig4}(b), is close to $\pm 1$ in wide regions of $\mu$
in the vicinity of the Dirac points. This is because the thermopower in the FM configuration is
then  negligibly small in comparison  to that in the AFM configuration. Thus, varying magnetic
configuration one can easily change not only  electrical resistance of the system, but
also the voltage generated by a temperature gradient.

\subsection{AP configuration}

\begin{figure}[ht!]
  \begin{center}
    \begin{tabular}{cc}
      \resizebox{70mm}{!}{\includegraphics[angle=0]{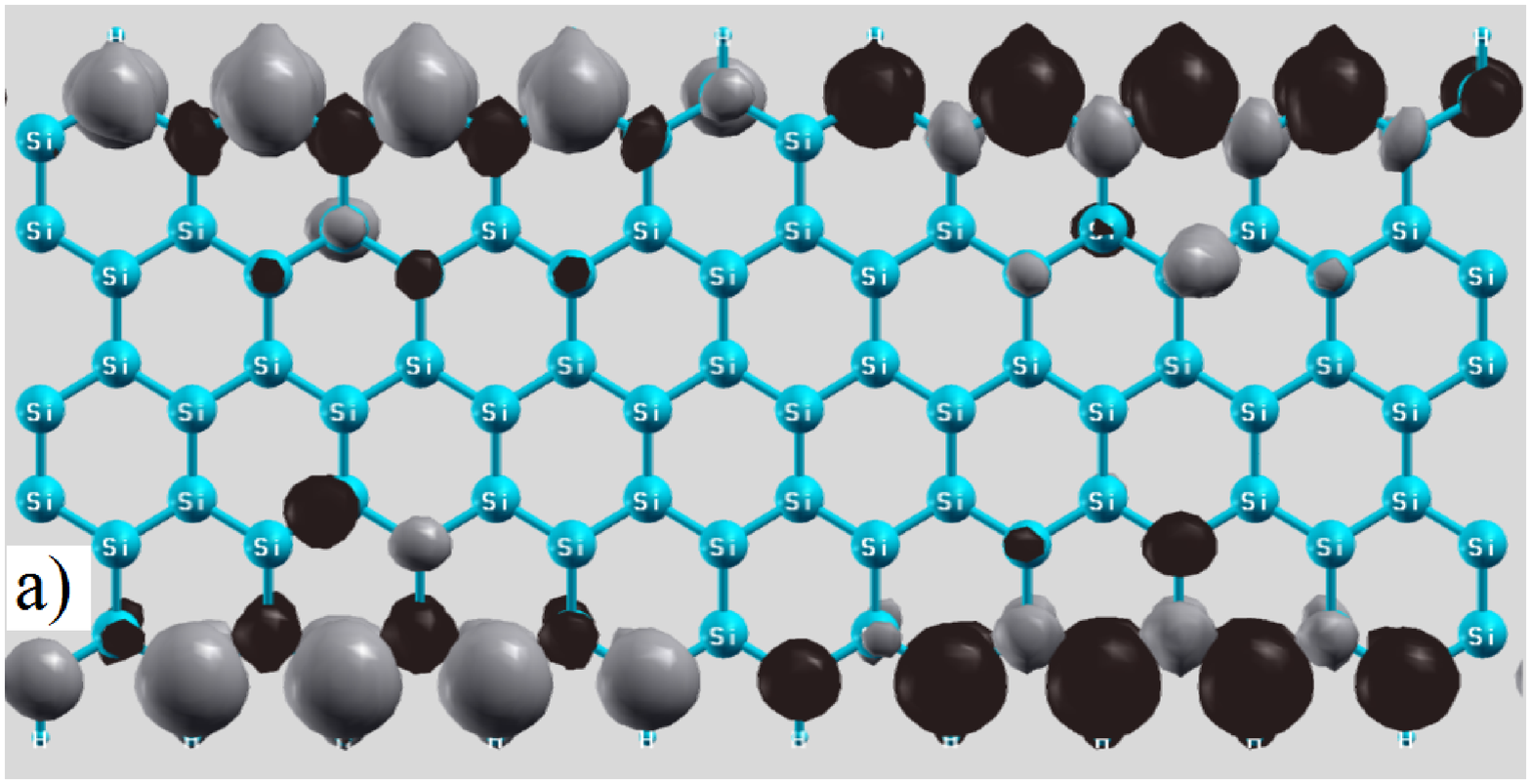}} \\
      \resizebox{80mm}{!}{\includegraphics[angle=270]{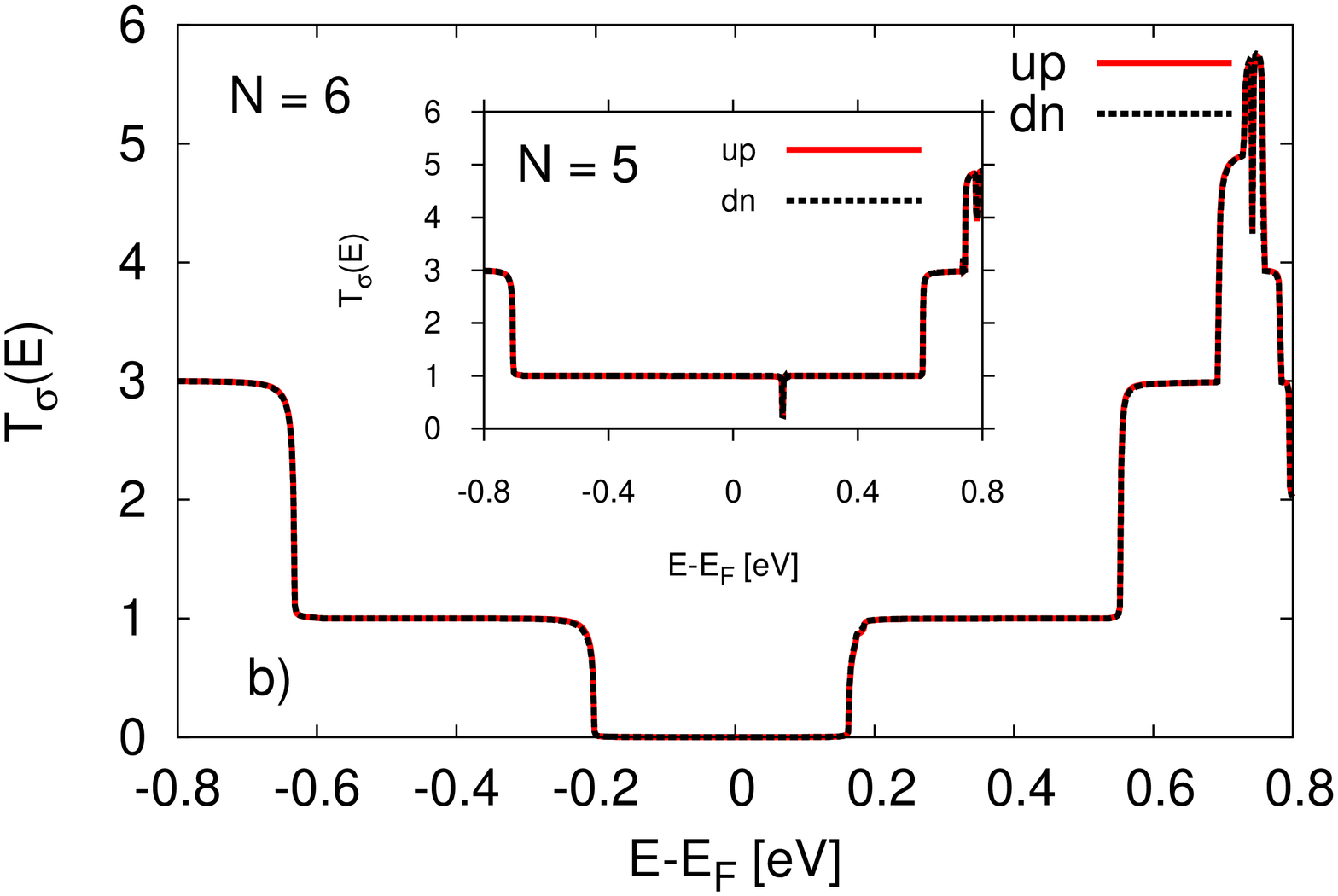}}
     \end{tabular}
    \caption{Spin density (a), and transmission $T_\sigma (E)$ for $N=6$ (b). The inset in (b) shows the transmission for $N=5$. Results obtained within the  GGA approximation.}
    \label{fig5}
  \end{center}
\end{figure}

\begin{figure*}[ht!]
  \begin{center}
    \begin{tabular}{cc}
      \resizebox{80mm}{!}{\includegraphics[angle=270]{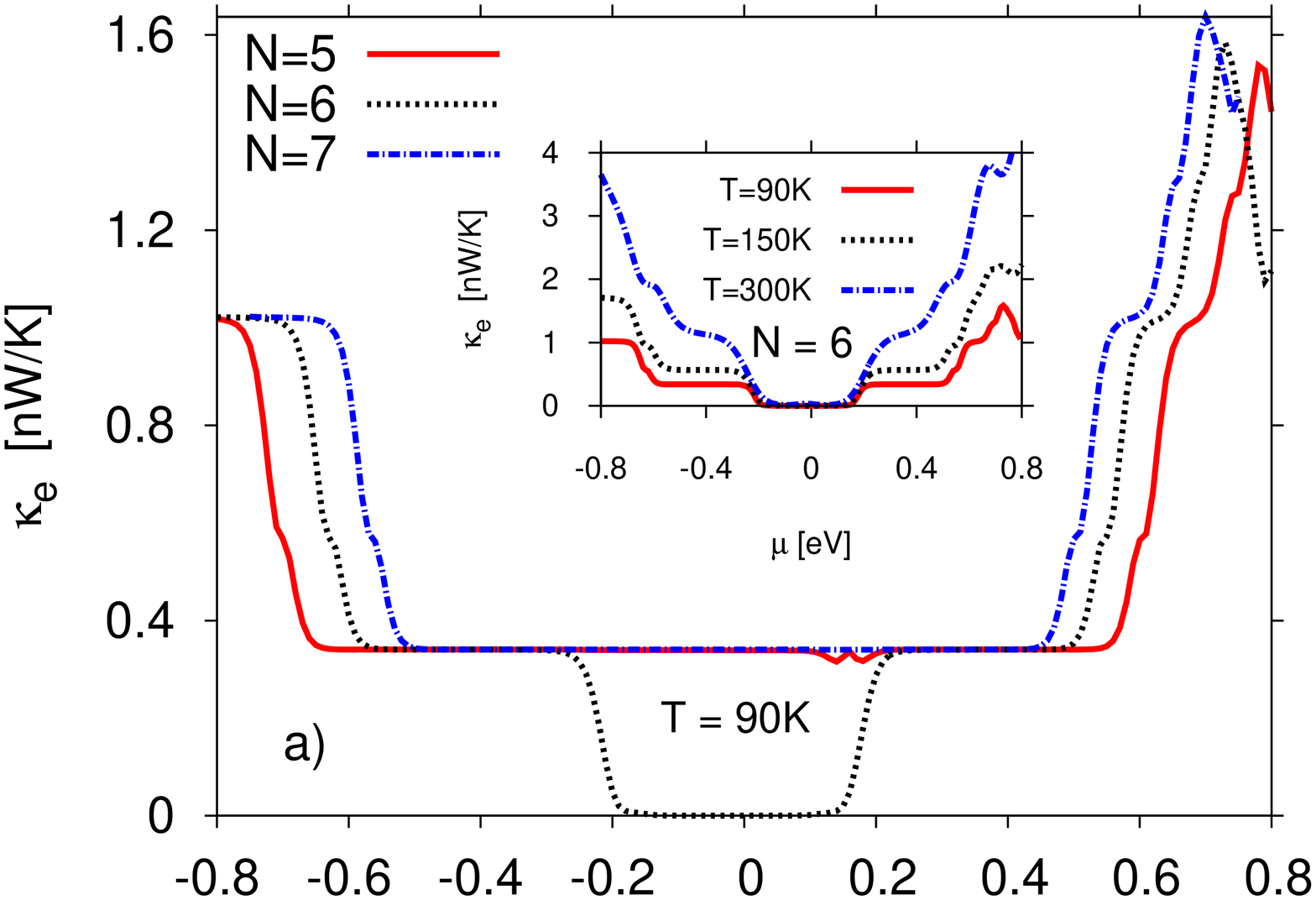}} &
      \resizebox{80mm}{!}{\includegraphics[angle=270]{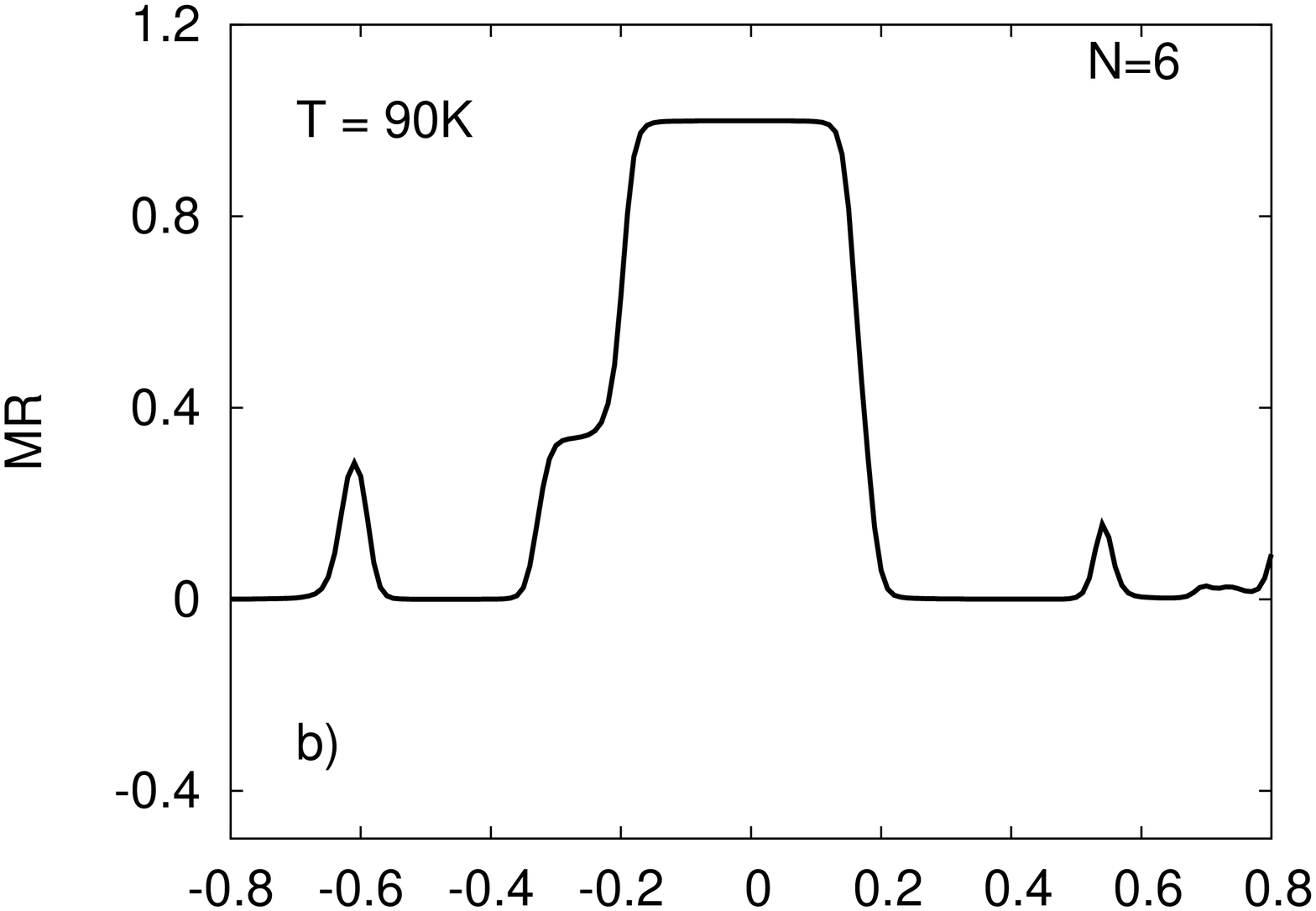}} \\
      \resizebox{80mm}{!}{\includegraphics[angle=270]{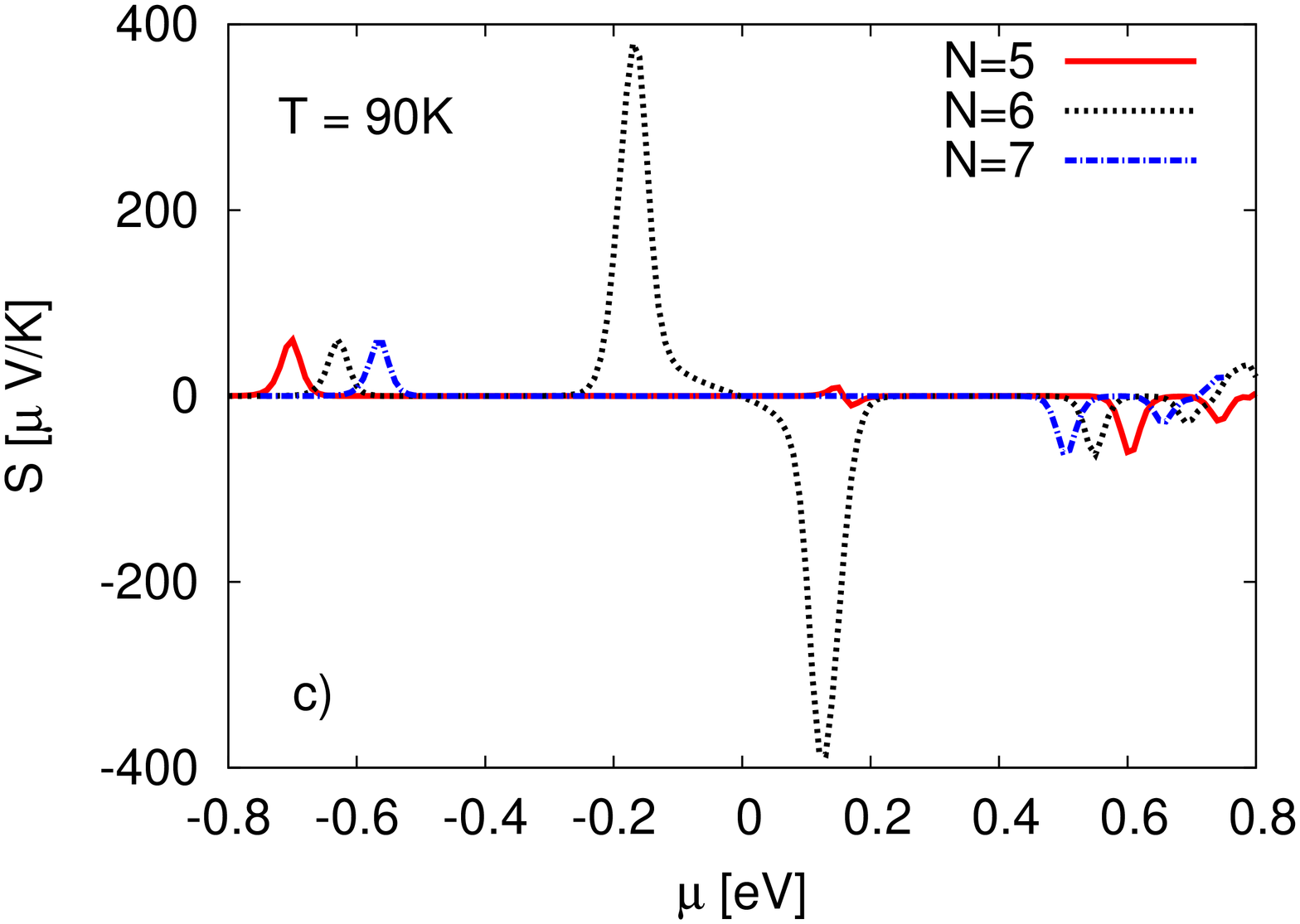}} &
      \resizebox{80mm}{!}{\includegraphics[angle=270]{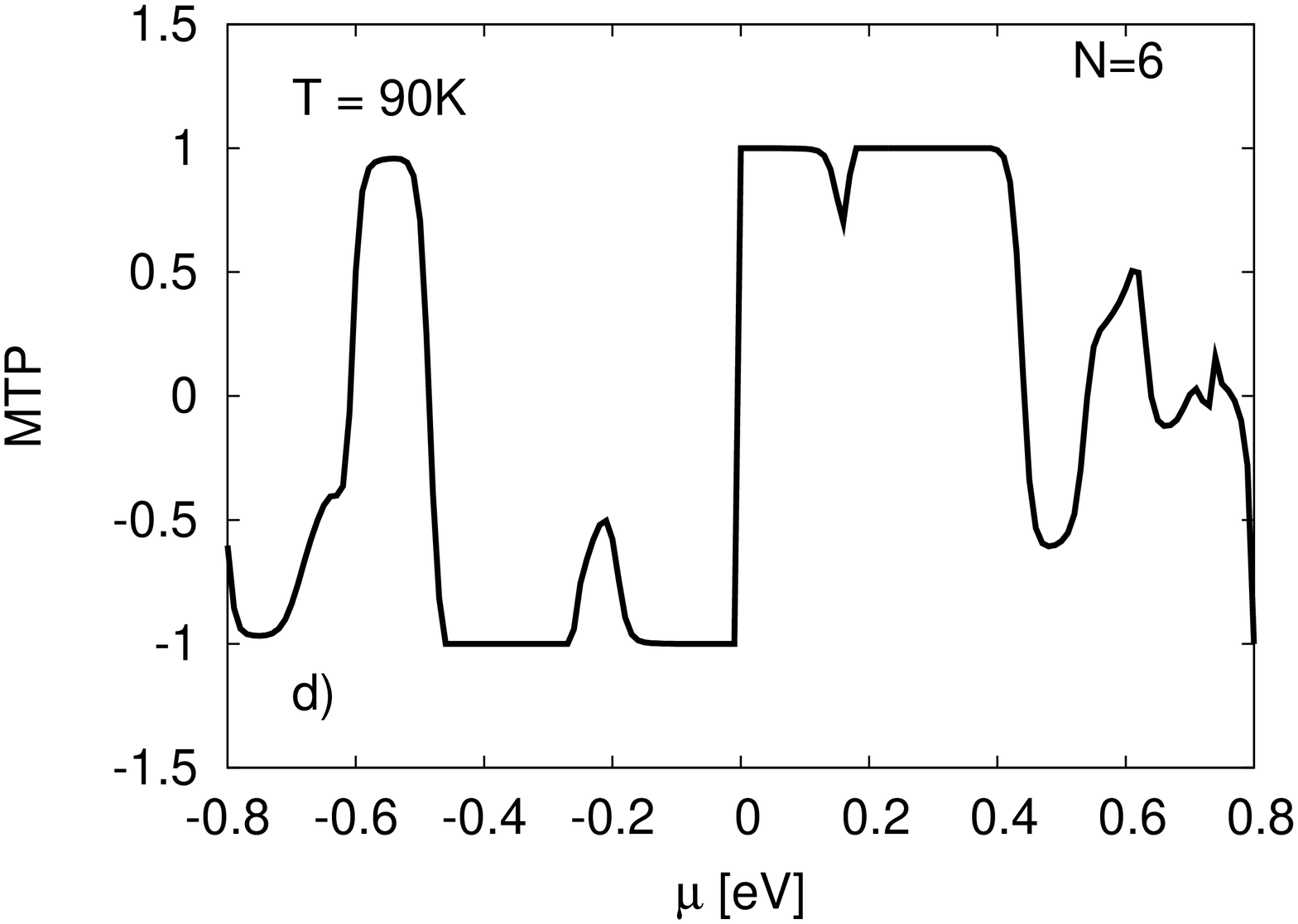}} \\
    \end{tabular}
    \caption{Electronic thermal conductance (a), and
             thermopower (c) as a function of $\mu$ for the  AP configuration, calculated within the GGA approximation.
             Parts (b) and (d) present MR and MTP  associated with transition from AP to FM configurations}
    \label{fig6}
  \end{center}
\end{figure*}

By applying a specific magnetic field, one can force the antiparallel (AP)
configuration, in which the spin polarization in the left part of the nanoribbon
is opposite to that in the right part. Thus, the AP state corresponds to the situation,
when the left and right electrodes are polarized antiparallel, as presented in Fig. \ref{fig5}(a).
Electronic transport in such a configuration strongly depends on the ribbon widths.
For $N$ odd, $N=5,7$, there is no gap in the band structure and the transmission is finite
and constant in the vicinity of $\mu=0$, while for $N=6$ a gap  opens near $\mu=0$, see Fig. \ref{fig5}(b). Note,
the band structure and transmission are now independent of spin orientation [10]. Thus, transport is strongly
suppressed in the nanoribbons with $N=6$, and the corresponding electrical and thermal
conductances are small as compared to those for $N=5,7$. This is shown in Fig. \ref{fig6}(a), where the thermal
conductance is presented for nanoribbons of different widths and at different temperatures. Note the vanishingly small conductance in the region close to $\mu=0$ for nanoribbons with $N=6$, and no such a suppression for other ribbons.
Due to the gap existing for $N=6$, there is a large change in the electric and heat conductances when  magnetic configuration
is changed from the AP to FM state. A large magnetoresistance  appears especially for chemical potentials close to the Dirac points, see Fig. \ref{fig6}(b). Much lower MR is expected for  for nanoribbons with odd $N$.

The thermopower in the AP configuration also strongly depends on the ribbon width.
Due to the presence of the energy gap for $N=6$, it is considerably enhanced in the vicinity of small $\mu$, similarly as in the case of AFM state, see Fig. \ref{fig6}(c). Accordingly, for ribbons with $N=6$, a
large MTP effect can be observed
when the configuration is changed from the AP state to the FM one, see Fig. \ref{fig6}(d).
Absolute value of MTP is close to unity in a wide range of chemical potential around $\mu=0$. On the other hand,
for zSiNRs with $N=5$ and $N=7$, the thermopower $S$ in the AP configuration is rather small in the
vicinity of $\mu=0$. All this demonstrates a large influence of the ribbon width on transport properties and thermopower in the AP
configuration, whereas  the ribbon width has
only a weak  influence on transport in the AFM and FM configurations.

\subsection{Comparison of the GGA and LDA methods}

The results of ab initio calculations presented in the previous sections were based
on the GGA method. These results clearly show that the energy gap, which appears in
the AFM and AP configurations, has a strong influence on the transport and thermoelectric coefficients
(see Figs \ref{fig2} and \ref{fig6}). It is also well known, that accurate determination of the energy
gap is the main problem in the DFT procedures, and the width of the gap can depend
on the used approximations. Therefore, it seems reasonable to compare the results
achieved within two main {\it ab initio} approximations, namely GGA and LDA. Here we limit
considerations to the AFM case, since the main conclusions also apply to the AP configuration with
an energy gap. On the other hand, the results for
configurations with constant and nonzero transmission in the vicinity of Fermi energy,
namely for the FM state and AP configuration with $N=5,7$, only weakly depend on the
used approximation.

\begin{figure*}[ht]
  \begin{center}
    \begin{tabular}{cc}
      \resizebox{85mm}{!}{\includegraphics[angle=270]{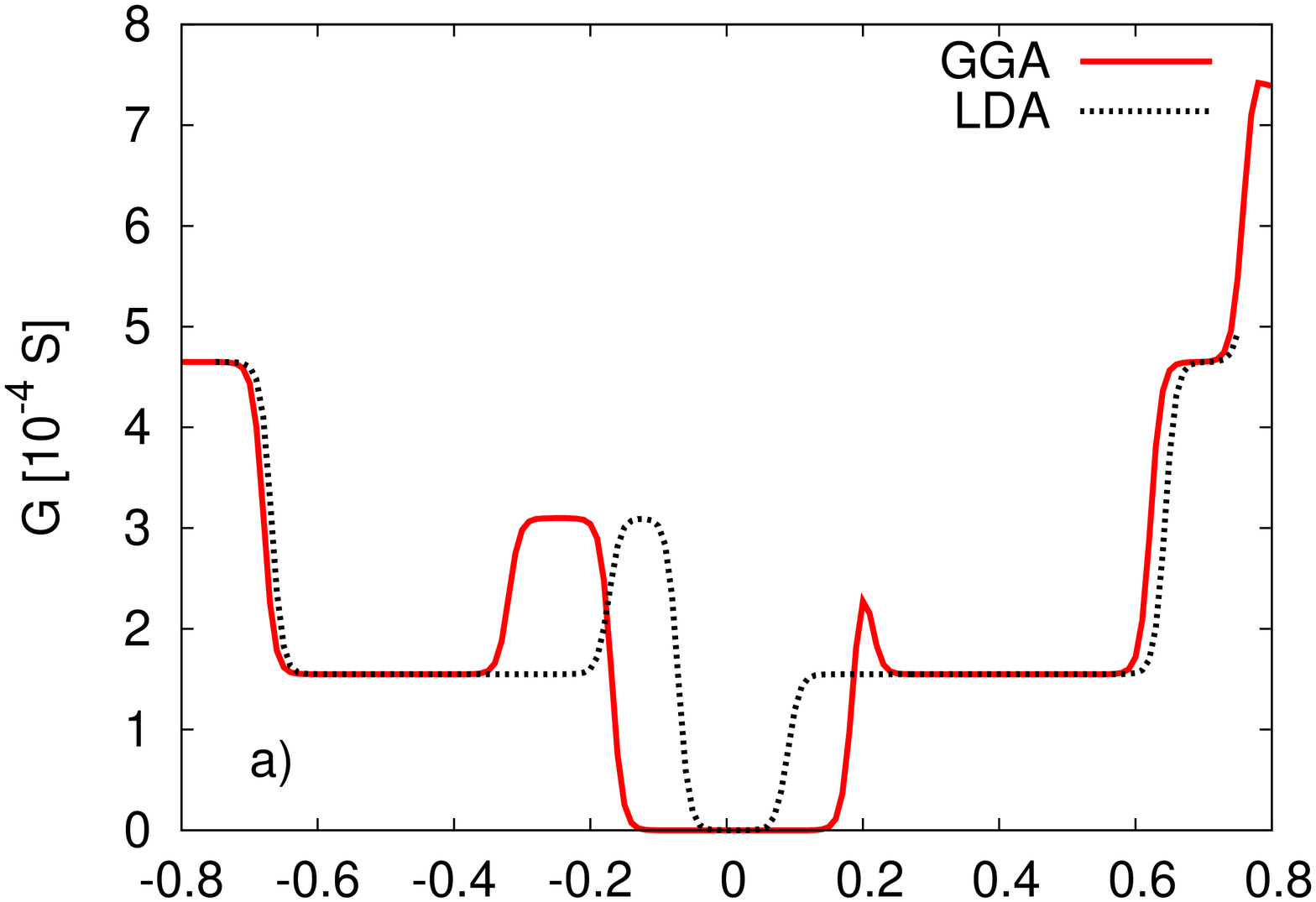}} &
      \resizebox{85mm}{!}{\includegraphics[angle=270]{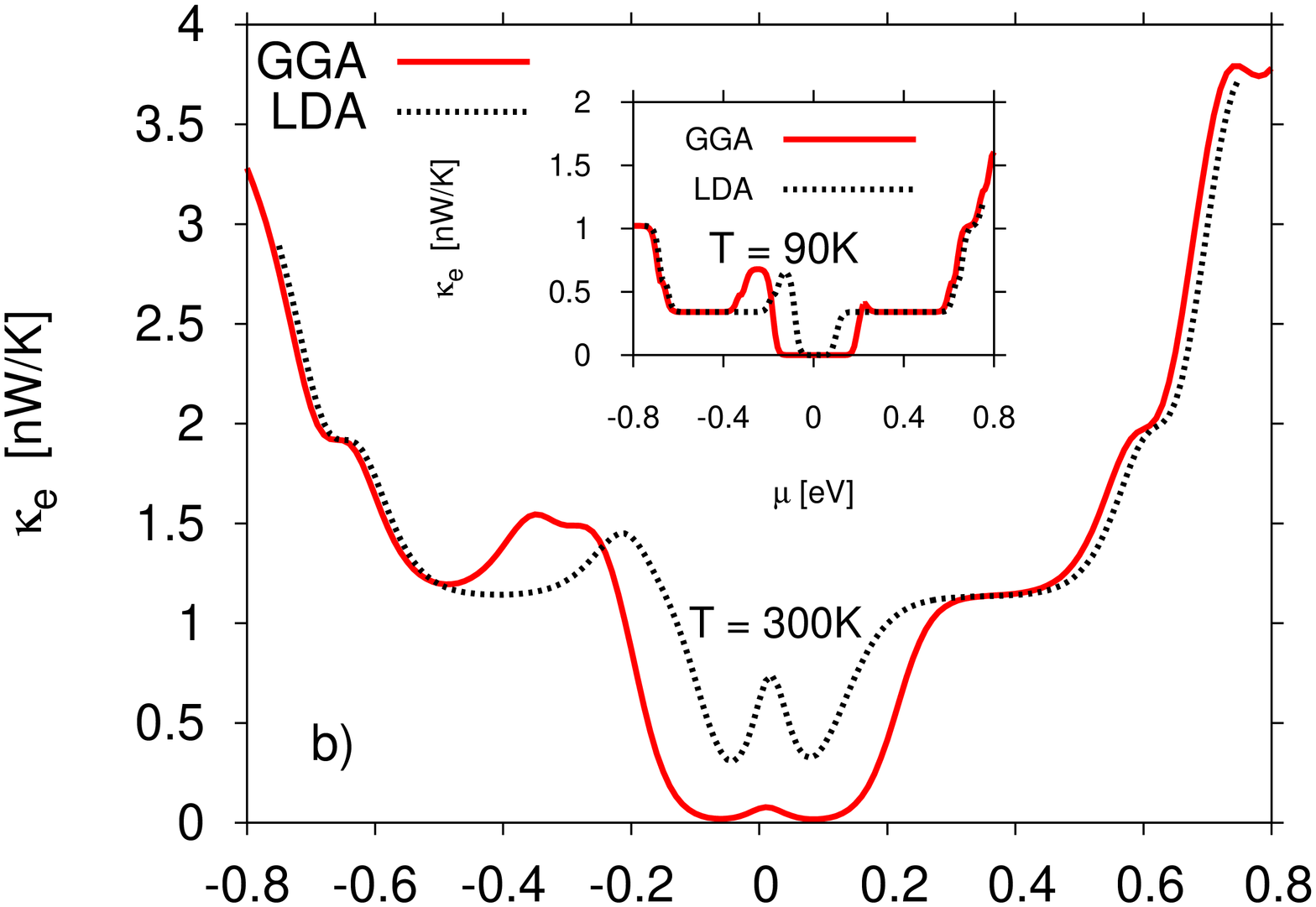}} \\
      \resizebox{85mm}{!}{\includegraphics[angle=270]{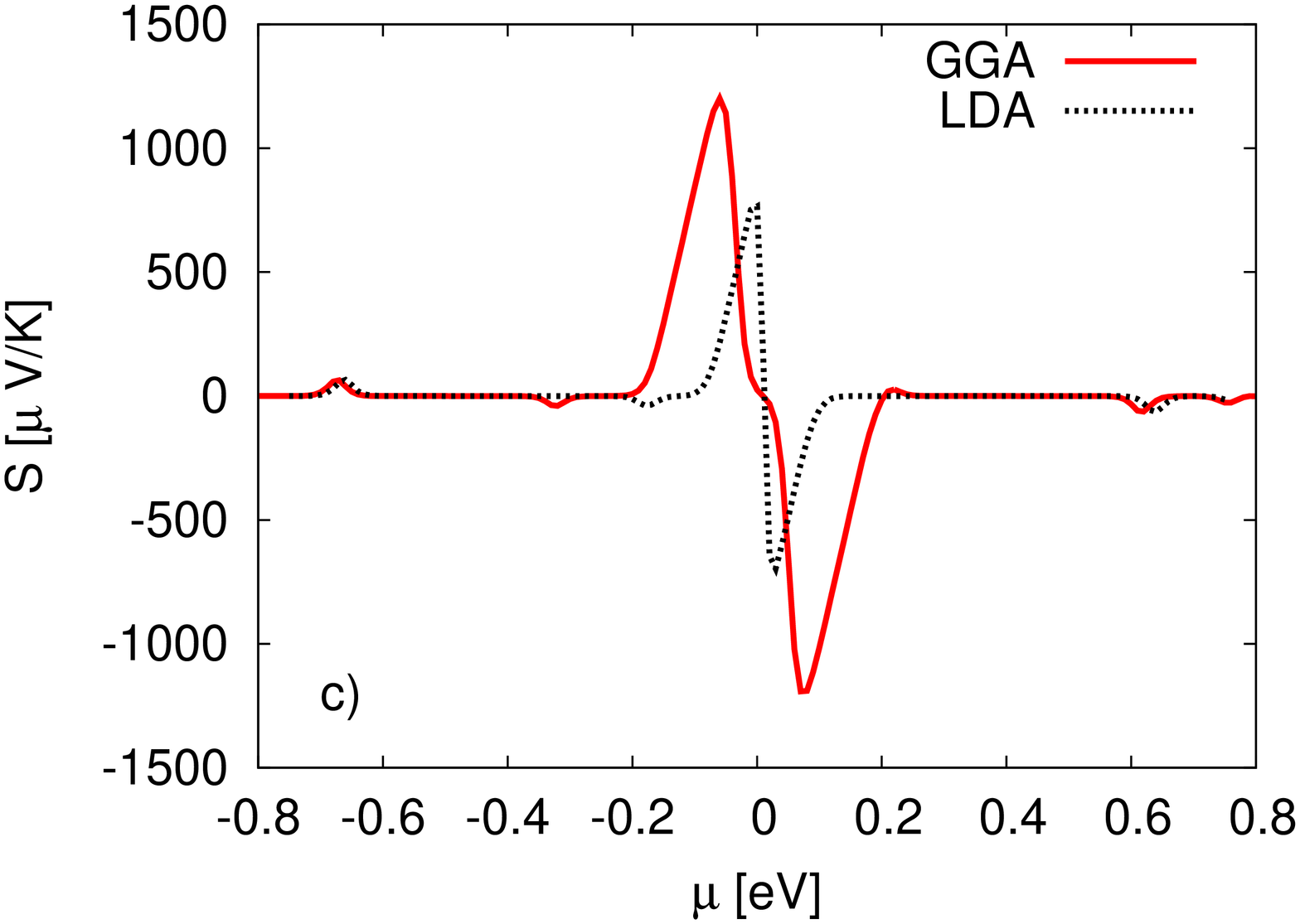}} &
      \resizebox{85mm}{!}{\includegraphics[angle=270]{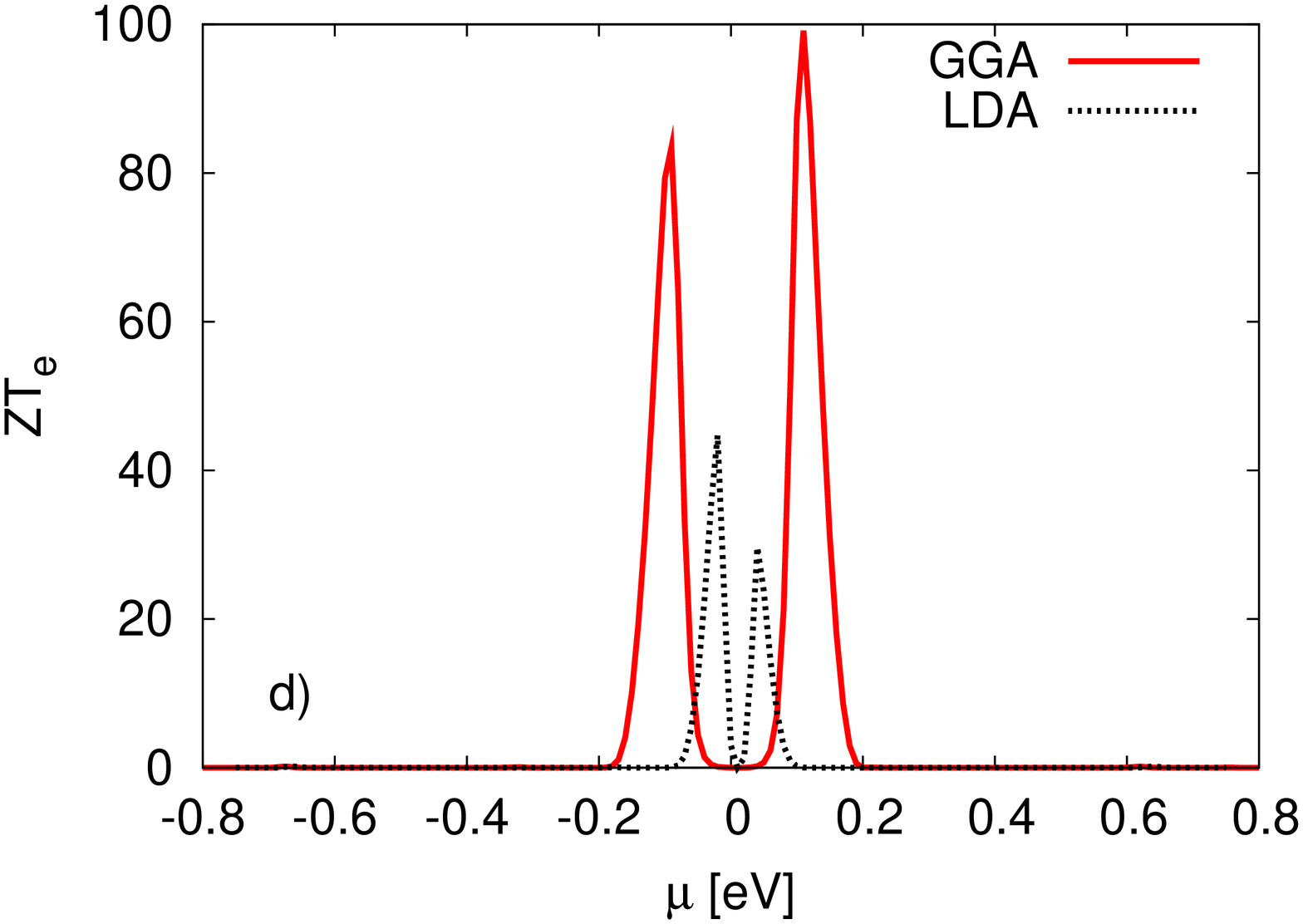}} \\
    \end{tabular}
    \caption{Electrical conductance $G$ (a), electronic contribution to the thermal conductance,
             $\kappa_{e}$ ( b ), thermopower $S$ (c), and figure of merit $ZT_{e}$ (d) as a function
             of chemical potential, calculated for the AFM configuration within the GGA (solid lines)
             and LDA (dashed lines) approaches for $N=5$ and $T=90$K}
    \label{fig7}
  \end{center}
\end{figure*}

In Fig. \ref{fig7} we compare the results obtained for the AFM configuration within
GGA and LDA methods. As a general rule, the energy gap calculated with the use of LDA
is remarkably narrower than for GGA. This, in turn, has a considerable influence on
the transmission function, leading to different behavior of the electrical and thermal
transport. Accordingly, at low temperatures, the electrical conductance $G$ and thermal
conductance $\kappa_{e}$, calculated within LDA method, are suppressed in a narrower
range of chemical potentials, see Fig.~\ref{fig7}(a,b).  These changes are even more pronounced at higher
temperatures, where $\kappa_{e}$(LDA) strongly increases revealing a quite remarkable
peak in the vicinity of small $\mu$. Results obtained for the thermopower $S$ also depend
on the calculation scheme. This is because the thermopower $S$ strongly depends on the gap
width. From Fig.~\ref{fig7}(c) follows that the dominant peaks in $S$ are suppressed and also shifted to
the region of smaller $\mu$ for the  LDA method. There is also some asymmetry in the peak
intensities for negative and positive $\mu$. The results calculated within LDA
approximation for higher temperatures are consistent with those presented by Pan et al.~\cite{pan}.
The strong asymmetry obtained in this reference is mainly due to the applied
approximations. In Fig. \ref{fig6}(d) we show the thermoelectric efficiency, $ZT_e=\frac{S^{2}GT}{\kappa_{e}}$.
In this formula the thermal conductance includes only electronic contribution. The
calculated $ZT_{e}$ is relatively high, mainly due to the strong suppression of electronic
thermal conductance. The influence of phonon contribution to the thermal conductance will
be discussed in the section 6. One can also see that the thermoelectric efficiency is
strongly dependent on the approach and for GGA is much higher than for LDA. Moreover, since the gap is  narrower in the LDA approximation, 
the corresponding peaks in $ZT_{e}$(LDA) are much closer to each other.

\section{Spin thermoelectric effects }

When the two spin channels are not mixed by spin-flip transitions and can be treated as independent in the
whole system, the temperature gradient can lead not only to charge accumulation
at the ends of an open system, but also to  spin accumulation. In other words,
the temperature gradient gives rise not only to  electrical voltage, but also
to spin voltage. To observe the spin voltage, the length of the sample should be
smaller than the spin-flip length. In the case under consideration, the system  consists of a nanoribbon
of length that is small enough, so the spin-flip scattering processes can be neglected.
In fact, the system is a part of a long nanoribbon, whose outer  parts form two external leads
(electrodes). Thus, spin thermoelectric properties can be observed, and we will focus
especially on the spin Seebeck effect. Moreover, the spin accumulation leads not only to the spin
Seebeck effect, but also modifies the conventional electrical thermopower (Seebeck effect)
as well as the electronic term in the heat conductance.

\begin{figure}[h]
    \begin{tabular}{c}
      \resizebox{85mm}{!}{\includegraphics[angle=270]{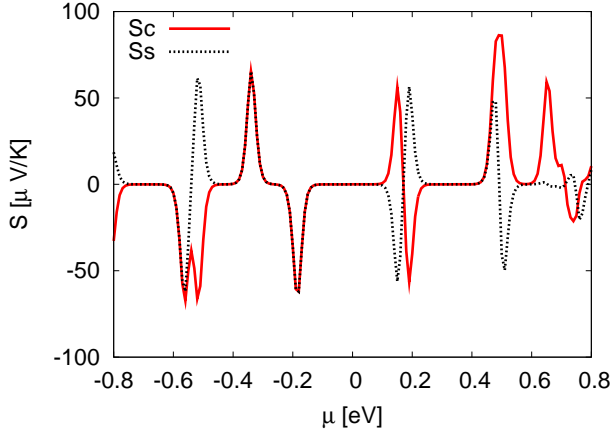}}
    \end{tabular}
    \caption{Charge (solid line) and spin (dashed line) thermopower as a function of chemical
             potential calculated for the FM configuration within GGA for $N=7$ and  $T=90$K}
    \label{fig8}
\end{figure}

When the two spin channels are independent,  one can introduce spin-dependent
thermopower, S$_{\sigma}$ ($\sigma=\uparrow,\downarrow$ ), defined as
$S_{\sigma}=-\frac{\Delta V_{\sigma}}{\Delta T} =-\frac{L_{1 \sigma}}{|e|TL_{0 \sigma}}$.
The spin-dependent thermopower corresponds to a spin-dependent voltage generated by a
temperature gradient~\cite{wierzbicki}. The spin-dependent moments $L_{n \sigma}$,
which appear in the above expression, are calculated with spin-dependent transmission
$T_{\sigma}(E)$. Generally, one can define then charge thermopower
$S_{c}=\frac{1}{2}(S_{\uparrow}+S_{\downarrow})$ as well as spin thermopower
$S_{s}=\frac{1}{2}(S_{\uparrow}-S_{\downarrow})$. Both $S_{c}$ and $S_{s}$, calculated
for the FM state at low temperature, are presented in Fig. \ref{fig8} as a function of $\mu$. Due to constant
transmission in the FM configuration, $S_{c}$ and $S_{s}$ are practically equal to zero
in a wide range of small $\mu$. For higher values of $\mu$, both $S_{c}$ and $S_{s}$ show several
peaks. It is interesting to note that intensities of the corresponding peaks in
$S_{c}$ and $S_{s}$ are similar. In systems with p-type doping, and in the vicinity of chemical
potential close to $-0.2$ eV, charge and spin thermopowers are practically equal and
change with $\mu$ in a similar way. On the other hand, for n-type doping in the vicinity
of $\mu \approx  0.2$ eV they have similar values but opposite signs. Such a behavior
is a result of the strong dependence of the transmission function on spin orientation. More specifically,
transmission for majority spins  shows a broad maximum for negative values of energy,
whereas it is constant for minority spins. Therefore, in this region practically there
is no contribution to $S_{c}$ and $S_{s}$ from minority spins and both quantities behave in a similar
way. On the other hand, for positive $\mu$  the main contribution comes from minority spins, which 
results in opposite signs of $S_{c}$ and $S_{s}$. The
charge thermopower in spin polarized systems is usually higher than the spin thermopower. It seems that the very peculiar behavior
of thermopower in zSiNRs is a unique feature of the system under consideration.

\section{Thermal conductance due to phonons}

The transport coefficients calculated above were limited to electronic contributions only.
However, an important contribution to heat current comes from phonons. In some specific
situations, especially when charge current is strongly suppressed, e.g. in the  energy gap, this contribution to heat current may be dominant.
This, in turn, can have a significant impact on thermoelectric efficiency which may be remarkably reduced as well. In this section we
present numerical results on the role of phonon term in the heat conductance in silicene. We will use two different methods to calculate the phonon dispersions, {\it ab-inito} and the model known as 4 Nearest-Neighbor Force Constants (4NNFC). Parameters of the latter model will be determined from fitting of the corresponding phonon spectrum to the one  from {\it ab-inito} calculations.

\subsection{Force-constants for silicene in the 4NNFC model}

The 4NNFC model, describing the dynamical matrix by 12 parameters
in the fourth-neighbor approximation,
was introduced by Saito \cite{saito} for carbon nanotubes, and subsequently
applied to graphene \cite{zimmerman}. Three force constant parameters are introduced for a given atom and its neighbor: $\phi_r$ for radial
displacement (bond-stretching), $\phi_{ti}$ for in-plane tangential displacement,
and $\phi_{to}$
for out-of-plane displacement. When the direction from a given atom to its
neighbor coincides with the $x$-axis, the resultant force-constant tensor
$K$ is diagonal:

\begin{equation}
K=
\left(
\begin{array}{ccc}
\phi_r & 0 & 0 \\
0 & \phi_{ti} & 0 \\
0 & 0 & \phi_{to}
\end{array}
\right).
\end{equation}

In a general case, when the direction from a given atom to its neighbor
is described by spherical angles $\phi$ and $\theta$,
one has to rotate the force-constant tensor by the following
orthogonal transformation:

$K'=R_x(\theta)^\dagger\,R_z(\phi)^\dagger\,K\,R_z(\phi)\,R_x(\theta)$,

where $R_z(\phi)$ is the matrix of rotation around the $z$-axis by $\phi$,
and $R_x(\theta)$ is the matrix of rotation around the $x$-axis by $\theta$.

We varied the 12 force-constant parameters to obtain the best fit
to ab-initio calculations. Since the structure of buckled silicene is not
flat, the dynamical matrix is not invariant under infinitesimal in-plane rotations.
Therefore, we do not apply the rotational invariance condition
for in-plane and out-of plane tangential force constants,
$\phi^{(1)}_t+6\phi^{(2)}_t+4\phi^{(3)}_t+14\phi^{(4)}_t=0$,
which is valid for graphene \cite{zimmerman}.
The {\it ab-initio} calculations were performed in 50 $k$-points along the high-symmetry
path in the first Brillouine zone of hexagonal two-dimensional lattice
with the use of Abinit code \cite{abinit1}. For calculation of the phonon spectra, the code
relies on the density-functional perturbation theory \cite{abinit2}.
The force-constants parameters were varied
to minimize the sum of squares of deviations from ab-initio values
of the phonon frequencies.
Table~\ref{tab1} presents the obtained force-constants,
which correspond to the global minimum.

\begin{table}
$$
\begin{array}{l|r|r|r}
\mbox{neigbour} & \phi_r & \phi_{ti} & \phi_{to}\\ \hline
\mbox{1st} & 2.0639 & 15.9965 & 0.3814 \\ \hline
\mbox{2nd} & -0.8961 & 0.9010 & 0.0683 \\ \hline
\mbox{3rd} & 0.2537 & -0.9737 & 0.1396 \\ \hline
\mbox{4th} & 0.3005 & -0.1067 & -0.1006 \\
\end{array}
$$
\caption{Force-constants for silicene in the 4NNFC model.\label{tab1}}
\end{table}

\begin{figure}[ht]
    \begin{tabular}{c}
      \resizebox{85mm}{!}{\includegraphics[angle=0]{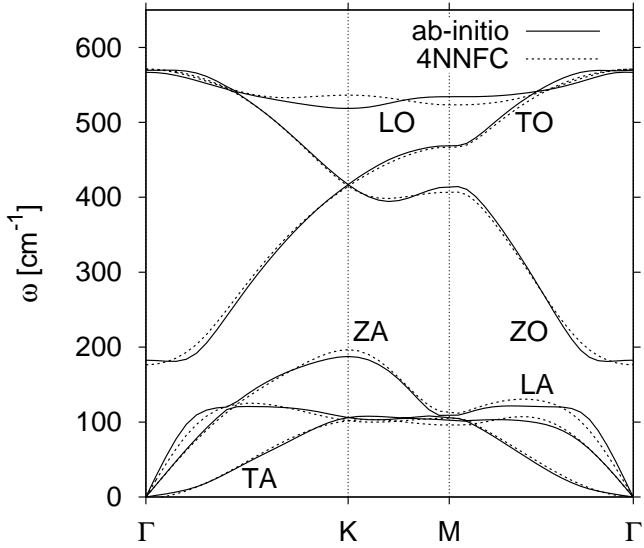}}
    \end{tabular}
    \caption{Phonon dispersion relation for silicene. Phonon modes denoted as in \cite{cheng}}
    \label{fig9}
\end{figure}

Figure~\ref{fig9} presents the comparison of phonon dispersion relations
calculated within the 4NNFC model with the corresponding {\it ab-initio} results.
In turn, Table \ref{tab2} contains comparison of the phonon frequencies at high
symmetry points from ab-initio calculations with those obtained by the 4NNFC model.
There are some quantitative differences, but qualitative agreement is satisfactory.

\begin{table}
$$
\begin{array}{l|l|r|r}
\mbox{point}	& \mbox{mode}	& \mbox{ab-initio}	& \mbox{4NNFC} \\ \hline
\Gamma		& \mbox{ZO}	& 183	& 176		\\
		& \mbox{TO/LO}	& 568	& 571		\\
K		& \mbox{TA/LA}	& 106	& 102		\\
		& \mbox{ZA}	& 187	& 188		\\
		& \mbox{TO/ZO}	& 417	& 415		\\
		& \mbox{LO}	& 519	& 536		\\
M		& \mbox{TA}	& 102	& 96		\\
		& \mbox{ZA}	& 105	& 105		\\ 		
		& \mbox{LA}	& 110	& 112		\\
		& \mbox{ZO}	& 414	& 407		\\
		& \mbox{TO}	& 469	& 467		\\
		& \mbox{LO}	& 534	& 523		\\
\end{array}
$$
\caption{Phonon frequencies in $\mbox{cm}^{-1}$ at high
symmetry points.\label{tab2}}
\end{table}

\subsection{Phonon conductance for silicene nanoribbons}

\begin{figure}[h]
    \begin{tabular}{c}
      \resizebox{85mm}{!}{\includegraphics[angle=0]{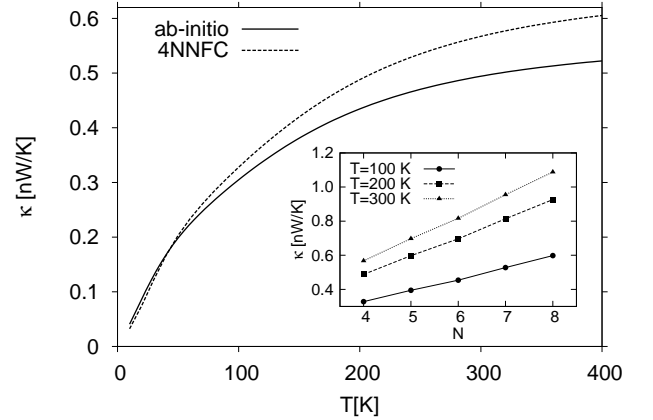}}
    \end{tabular}
    \caption{Phonon contribution to heat conductance of zSiNRs corresponding to $N=4$  as a function
             of temperature,  calculated  within the 4NNFC  and {\it ab-initio} models. The inset shows the phonon contribution to heat
             conductance calculated with the 4NNFC model as a function of N (zSiNRs width).}
    \label{fig10}
\end{figure}

Next, the obtained force-constants  are used to determine the dispersion relations for zigzag silicene nanoribbons within the 4NNFC model. From these dispersion relations one can determine the phonons transmission function T($\omega$), which is equal to the number of bands at the phonon energy $\omega$. \newline
\indent Phonon contribution to the heat conductance, $\kappa_{ph}$, can be determined by integration
of the transmission function, according to the following formula:
\begin{equation}
\kappa_{ph}=\frac{1}{2\pi}\int\limits_0^\infty
\hbar\omega\,T(\omega)\,\frac{\partial n}{\partial T}\,
d\omega
\end{equation}
where $n$ is the Bose-Einstein distribution function of equilibrium phonons
at temperature $T$. The phonon contribution to heat conductance  for zSiNRs with $N=4$
is presented in Fig.~\ref{fig10} as a function of temperature.
The inset to Fig.~\ref{fig10} shows the dependence of the phonon conductance within the 4NNFC model on the width
of zSiNRs (the points correspond to $\mbox{N}=4,5,6,7,8$). As expected, $\kappa_{ph}$, depends linearly
on the nanoribbon width. This phonon term in heat conductance is compared there with that obtained from {\it ab initio} method.
The ab-initio transmission function was obtained from phonon spectrum calculated
with use of phonopy code \cite{phonopy}, which realizes
the Parlinski-Li-Kawazoe method, based on the supercell approach with the finite
displacement method \cite{parlinski}. Forces acting on atoms with respect to their
displacements, needed by this method, were calculated with the VASP code \cite{vasp1,vasp2}
using PAW pseudopotentials \cite{vasp3}. The difference between the results obtained within the two approaches is minor at low temperatures, but grows as the temperature increases.

\subsection{Influence of phonons on thermoelectrical efficiency}

\begin{figure}[ht]
  \begin{center}
    \begin{tabular}{cc}
      \resizebox{80mm}{!}{\includegraphics[angle=270]{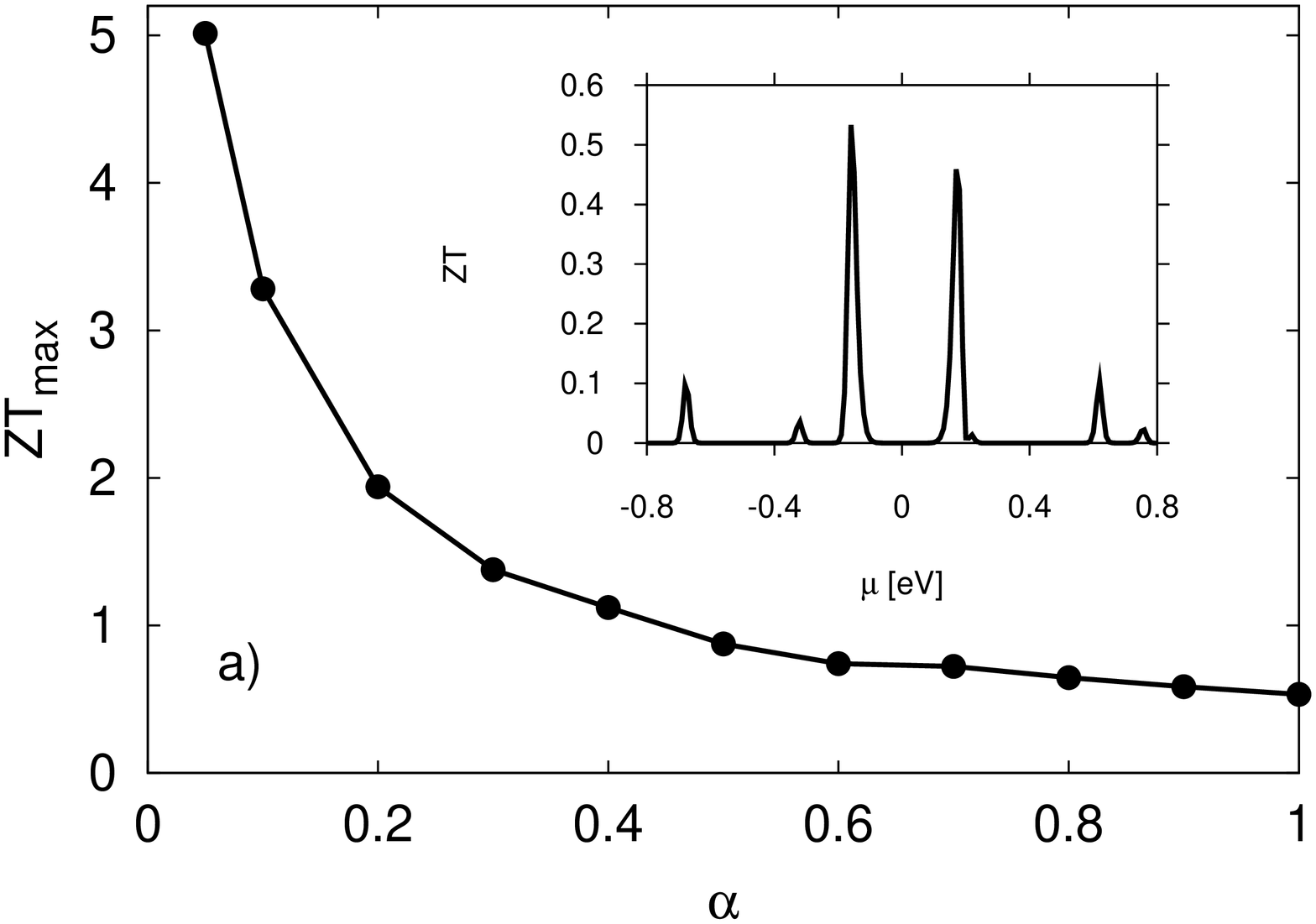}}\\ 
      \resizebox{85mm}{!}{\includegraphics[angle=270]{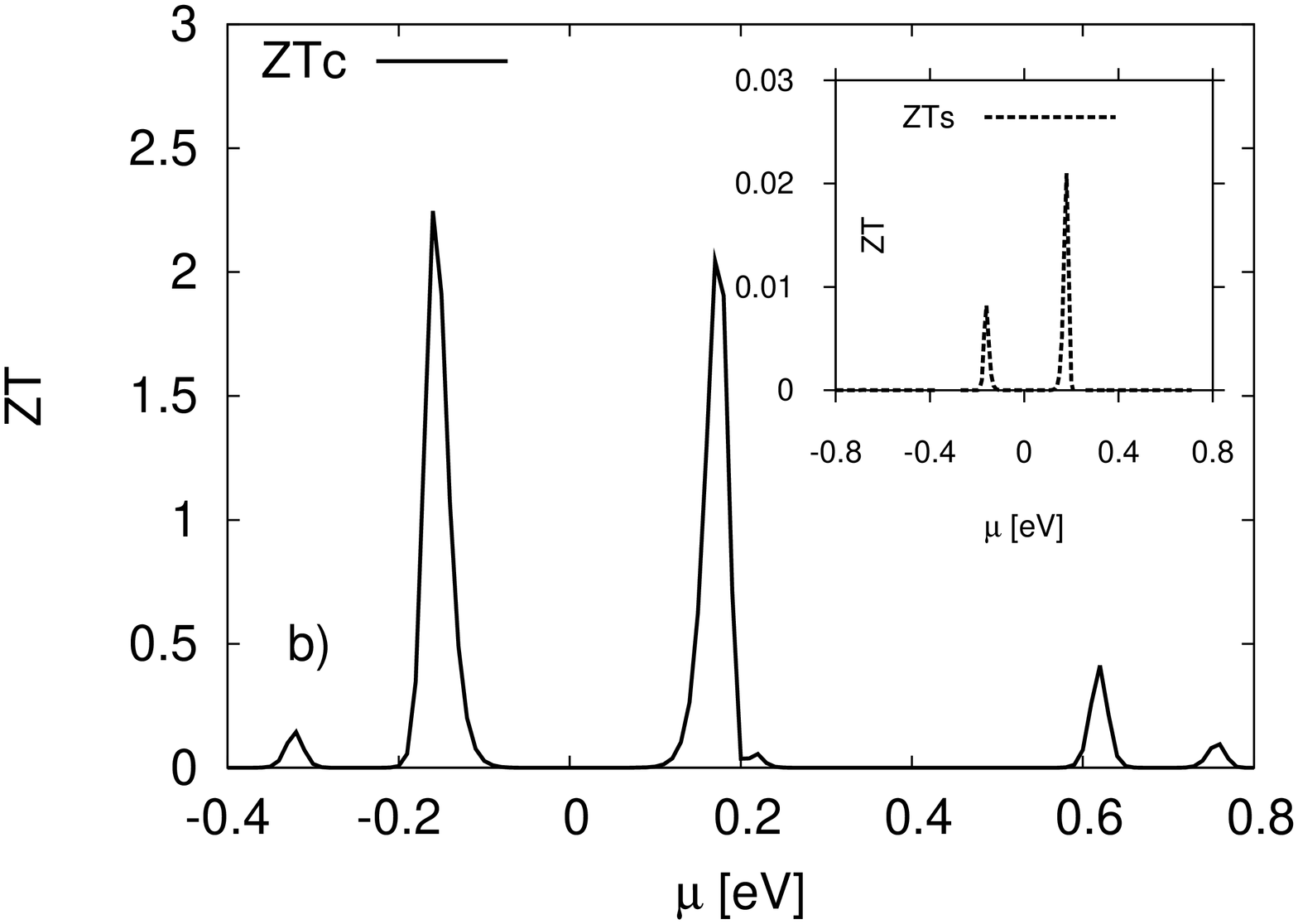}}
    \end{tabular}
    \caption{Maximum of $ZT$ for AFM case as a function of scaling parameter $\alpha$ (a).
             $ZT$ for AFM case as a function of chemical potential (inset to fig a),
             $ZT_{c}$ ($ZT_{s}$ in inset) for FM configuration (b). $T=90$ K, $N=5$, GGA approach}
    \label{fig11}
  \end{center}
\end{figure}

The influence of phonons on thermoelectric efficiency $ZT$ is presented in Fig.~\ref{fig11}(a)  for AFM case (inset).
The calculated phonon conductance strongly suppresses $ZT$, which
appears to be lower than 1 in the whole region under consideration.
We note that the authors of~\cite{pan}, using molecular
dynamic simulations, obtained relatively low phonon conductance, even at
high temperatures. This shows that the accurate determination
of $\kappa_{ph}$ in silicene nanoribbons is a difficult task. Moreover, the
interaction between the narrow nanoribbons and a substrate as well as electron-phonon
coupling can be important and may influence the phonon conductance. One can expect that
interaction with a substrate will considerably reduce $\kappa_{ph}$. To take into
account the reduction of thermal conductance due to a substrate, as well
as some differences in estimation of the phonon conductance when using different calculation methods,
we scale $\kappa_{ph}$ with a parameter $\alpha$, similarly as in Ref.~\cite{gunst},
namely $\kappa_{ph}$ determined in the previous subsection is expressed in the form:
$\alpha \kappa_{ph}$  and we discuss the influence of the parameter $\alpha$ on the
maximum value of $ZT$. According to Fig.~\ref{fig11}(a), one can see that $\alpha$ strongly affects 
the efficiency. The main maximum in $ZT$ is considerably suppressed even for $\alpha=$0.05.
For higher values of $\alpha$ the changes are not so rapid, but $ZT$ becomes relatively
low.

 It is interesting that relatively high efficiency
$ZT_{c}=S_{c}^{2}(G_{\uparrow}+G_{\downarrow})T/((\kappa_{e \uparrow}+\kappa_{e \downarrow})+\kappa_{ph})$ can be obtained
for FM configuration when the two spin channels are not mixed and spin effects are
important. As presented in the Fig.~\ref{fig11}(b)  maximum efficiency is as high as 2.5 despite
considerable phonon conductance. On the other hand, the spin part
$ZT_{s}=S_{s}^{2}(G_{\uparrow}-G_{\downarrow})T/((\kappa_{e \uparrow}+\kappa_{e \downarrow})+\kappa_{ph})$
is considerably suppressed.

\section{Summary and conclusions}

We have carried out detailed analysis of linear thermoelectric effects in silicene zigzag nanoribbons. Such nanoribbons may reveal several stable magnetic configurations of the edge magnetic moments: AFM (magnetic moments at one edge are opposite to those at the other edge), FM (magnetic moments at the two edges are align in parallel), and AP (magnetic moments in left part of the nanoribbon are opposite to those in the right part). The former configuration (AFM) is of the lowest energy, but the other two can be stabilized by an external magnetic field. Transmission function reveals a relatively wide gap at the Fermi level in the case of AFM state, while no gap appears in the FM state.

Thermoelectric parameters have been calculated in two limiting situations; (i) no spin accumulation can build up in the leads, and (ii) spin accumulation can appear due to slow or absence of spin relaxation. In the latter case spin thermoelectric effects can occur, especially the spin thermopower, which effectively describes spin voltage generated by a temperature gradient.
Electronic contributions to the thermoelectric effects reflect the presence of the gap, where the thermopower is significantly enhanced.
We have also calculated the phonon contribution to heat conductance, and thus also to the thermoelectric efficiency.
The phonon term in the heat conductance is dominant when the Fermi level is inside the energy gap, while both electronic and phonon contributions are comparable for Fermi level outside the gap. Thus, the phonon contribution suppresses the high value of the thermoelectric efficiency as well as of the spin thermoelectric efficiency, which were obtained for the Fermi levels inside the energy gap and when only electronic term in the heat conductance was taken into account.

For calculating the electronic and phonon transmission functions, we used various numerical procedures and approximations. There are some quantitative differences in the results obtained with those methods,  especially when the gap appears at the Fermi level of the nanoribbons. However, there is a good qualitative agreement between the results obtained with different methods.

\begin{acknowledgments}
This work was supported by the National Science
Center in Poland as the Project No. DEC-2012/04/A/ST3/00372.
Numerical calculations were performed at the Interdisciplinary Centre for
Mathematical and Computational Modelling (ICM) at Warsaw University and partly
at SPINLAB computing facility at Adam Mickiewicz University.
\end{acknowledgments}

\end{document}